# Chelyabinsk – a rock with many different (stony) faces: An infrared study

Andreas Morlok **(Corresponding Author)**, Institut für Planetologie, Westfälische Wilhelms-Universität Münster, Wilhelm-Klemm-Str. 10, 48149 Münster, Germany; Email: morlokan@uni-muenster.de; Telephone: +49 0251 8339069

Addi Bischoff, Institut für Planetologie, Westfälische Wilhelms-Universität Münster, Wilhelm-Klemm-Str. 10, 48149 Münster, Germany; Email: bischoa@uni-muenster.de

Markus Patzek, Institut für Planetologie, Westfälische Wilhelms-Universität Münster, Wilhelm-Klemm-Str. 10, 48149 Münster, Germany; Email: markus.patzek@uni-muenster.de

Martin Sohn; Hochschule Emden/Leer, Constantiaplatz 4, 26723 Emden, Germany; Email: martin.sohn@hs_emden-leer.de

Harald Hiesinger, Institut für Planetologie, Westfälische Wilhelms-Universität Münster, Wilhelm-Klemm-Str. 10, 48149 Münster, Germany; Email: hiesinger@uni-muenster.de

**Abstract**

In order to provide spectral ground truth data for remote sensing applications, we have measured mid-infrared spectra (2-18 µm) of three typical, well-defined lithologies from the Chelyabinsk meteorite that fell on February 15, 2013, near the city of Chelyabinsk, southern Urals, Russia. These lithologies are classified as (a) moderately shocked, light lithology, (b) shock-darkened lithology, and (c) impact melt lithology. Analyses were made from bulk material in four size fractions (0-25 µm, 25-63 µm, 63-125 µm, and 125-250 µm), and from additional thin sections.

Characteristic infrared features in the powdered bulk material of the moderately shocked, light lithology, dominated by olivine, pyroxene and feldspathic glass, are a Christiansen feature (CF) between 8.5 and 8.8 µm; a transparency feature (TF) in the finest size fraction at ~13 µm, and strong reststrahlen bands (RB) at ~9.1 µm, 9.5 µm, 10.3 µm, 10.8 µm, 11.2-11.3 µm, 12 µm, and between 16 and 17 µm. The ranges of spectral features for the micro-FTIR spots show a wider range than those obtained in diffuse reflectance, but are generally similar.

With increasing influence of impact shock from 'pristine' LL5 (or LL6) material (which have a low or moderate degree of shock) to the shock-darkened lithology and the impact melt lithology as endmembers, we observe the fading/disappearing of spectral features. Most prominent is the loss of a 'twin peak' feature between 10.8 and 11.3 µm, which turns into a single peak. In addition, in the 'pure' impact melt "endmember lithology" features at ~9.6 µm and ~9.1 µm are also lost. These losses are most likely correlated with decreasing amounts of crystal structure as the degree of shock melting increases. These changes could connect mid-infrared features with stages for shock metamorphism (Stöffler et al., 1991): Changes up to shock stage S4 would be minor, the shock darkened lithology could represent S5 and the impact melt lithology S6 and higher.

Similarities of the Chelyabinsk spectra to those of other LL chondrites indicate that the findings of this study could be related to this group of meteorites in general.

## 1. Introduction

The Chelyabinsk meteorite fell on February 15, 2013, near the city of Chelyabinsk, southern Urals, Russia. At least 1 t of material was collected from the strewn field, with the main mass consisting of a ~540 kg piece collected from Chebarkul Lake (Ruzicka et al., 2015).

The mass and pristine nature of the sample, as well as the spectacular nature of the fall, motivated researchers to extensively study the properties of the Chelyabinsk meteorite. The meteorite was officially classified as LL5 (e.g., Galimov, 2013, Ruzicka et al., 2015), but it was soon recognized as a breccia having different lithologies mixed together (Bischoff et al., 2013). Chelyabinsk is polymict only considering the different lithologies, but they are exclusively related to the LL chondrite parent body, indicating it is a genomict, LL5-6 breccia (Bischoff et al., 2006, 2013). The meteorite was thoroughly characterized in several consortium studies and various other investigations. Popova et al. (2013), Bezaeva et al. (2014), Kohout et al. (2014), Maksimova et al. (2014), Pavlov et al. (2014), Nabelek et al. (2015), Ozawa et al. (2015), Povinec et al. (2015), Reddy et al. (2015), and Righter et al. (2015) have studied the meteorite's properties using techniques including scanning electron microscopy (SEM), electron microprobe analysis (EMPA), transmission electron microscopy (TEM), X-ray diffraction, spectroscopy in the visible and near-infrared range, Raman spectroscopy, X-ray tomography, magnetic susceptibility, scanning magnetic microscopy, thermoluminescence, photoluminescence, Mössbauer spectroscopy, FTIR mass spectrometry of organic material, and isotope analyses of O, Cr, Sm, Cr, Ca, noble gases, U-Pb, Ar-Ar, Sm-Nd, and cosmogenic radionuclides.

Thanks to the fast recovery of the sample materials, there are only weak signs of terrestrial contamination and oxidation (Righter et al., 2015). While the degree of shock metamorphism overall is moderate (S4; Stöffler et al., 1991; Bischoff and Stöffler, 1992), the sample shows a very complex collision history, with a minimum of 8 impact events between ~4.53 Ga and 27 Ma (Righter et al., 2015). There are three major lithologies recognizable as a result of the shock metamorphism. Moderately-shocked LL5 - LL6 material is recognized by its light appearance in optical microscopy and in examining the specimen by hand. Shock-darkened material is the second typical lithology, while the third type can be characterized as an impact melt lithology (e.g., Bischoff et al., 2013; Kohout et al., 2014; Povinec et al., 2015; Righter et al., 2015). Shock veins were identified in nearly all lithologies (light-colored and shock-darkened fragments), but do not occur as distinct veins in the impact melt breccia fragments (Bischoff et al., 2013).

Previous spectroscopic studies of Chelyabinsk have focused on the visible and near-infrared (Vis/NIR) range commonly used for asteroid remote sensing.

The LL5 lithology and bulk meteorite material show typical Vis/NIR spectral features (Popova et al., 2013; Kohout et al., 2014), while the shock-darkened and impact melt lithologies exhibit very weak silicate features and much lower reflectance (Kohout et al., 2014; Reddy et al., 2015). By extending the wavelength range through analytical studies using mid-infrared spectral techniques, we are able to obtain additional spectral features for better discrimination of the existing lithologies.

Some earlier studies have investigated mid-infrared reflectance data for carbonaceous chondrites (McAdam et al., 2015) and enstatite chondrites (Izawa et al., 2010), but comparatively fewer studies have examined the reflectance and emission of ordinary chondrites in the mid-infrared range. A comprehensive study on a large range of chondrites and achondrite samples based on extensively characterized material (Jarosewich, 1990) was done by Salisbury et al. (1991b)—the data is available in the NASA RELAB (Reflectance Experiment Laboratory) Database at Brown University (Pieters, 1983; Miliken et al., 2016). Additionally, a series of spectra obtained from 0-45 µm size fractions of powdered and sieved bulk meteorites is available at the Johns Hopkins University Spectral Library (Baldridge et al., 2009). Miyamoto and Zolensky (1994) and Sato et al. (1997) studied volatile features in CM, CR, and CI chondrites in the mid-infrared. Additional transmission analyses of ordinary chondrites in the mid-infrared range were obtained by Sandford (1984) and Morlok et al. (2014), but these cannot be directly related to reflectance studies.

Similar to mid-infrared spectral analyses, micro-FTIR (Fourier transformation infrared spectroscopy) reflectance studies of chondrites are also still relatively rare. However, they have become increasingly popular in recent years, as recent technical advances have made FTIR data an increasingly useful and powerful technology for the fast in-situ microscopic study of meteorites. Recently, Hamilton et al. (2016) used micro-FTIR to map the mineralogy of a series of 24 carbonaceous chondrites. Further micro-FTIR studies were made of lithologies in CM chondrites by Djouadi et al. (2012) and organics in the Tagish Lake meteorite were studied by Kebukawa et al. (2005). The petrology of CAIs and chondrules were analyzed by Morlok et al. (2004a and b) and Hamilton and Connolly Jr. (2012). Benedix et al. (2016) studied Martian meteorites using this technique. Overall, micro-FTIR data provide an additional means for the mineralogical characterization of meteorite samples on a small scale.

Reflectance micro-FTIR is also used for specific studies of shock effects. For example, Jaret et al. (2015) analyzed experimentally shocked feldspars in basalts from the Lonar Crater; Martin et al. (2016) studied the shock history of lunar meteorites by distinguishing different types of shocked and unshocked materials; and Stephen et al. (2014) focused on shocked mineral phases in Martian meteorites. Thus, while using micro-FTIR to examine shock effects is still new, it is poised to become a powerful tool for investigating this property of meteorites

This study investigated the three different shock metamorphism-based lithologies of Chelyabinsk with the aims of (a) providing mid-infrared diffuse reflectance spectra of different grain size fractions for the representative lithologies in the Chelyabinsk meteorite and (b) obtaining in-situ micro-FTIR data from polished thin sections to provide a detailed picture of the spectral properties within the samples. These infrared data can be applied for remote-sensing purposes of asteroids, with the main goal in helping to identify potential parent bodies of Chelyabinsk. Candidates for the origin of the Chelyabinsk meteorite are Flora or Batistina families, and Near Earth asteroid 2011 EO40 (de la Fuente et al. 2013; Reddy et al., 2014; Reddy et al., 2015).

Also of interest is the possibility to link spectral features in the infrared to the degrees of shock metamorphism. Stöffler et al. (1991) proposed six stages of shock metamorphism for ordinary chondrites based on petrographic shock features in olivine and plagioclase. Ordinary chondrites that underwent weak impact shock (shock stages S2 and S3) up to 15-20 GPa and post-shock temperatures of up to 150°C show mainly planar fractures and undulatory extinction of the minerals, with only local melt formation. Moderately shocked material (S4; up to 35 GPa and 350°C) shows weak mosaicism and may contain interconnecting melt veins. At stage S5 (up to 55 GPa and 850°C) olivine shows strong mosaicism and most plagioclase has been transformed to maskelynite. These rocks may also contain abundant melt pockets, veins, and dykes (Stöffler et al., 1991). Very strongly shocked ordinary chondrites (S6) feature solid state recrystallization of olivine, ringwoodite formation, and shock melting of plagioclase. At pressures over 75-90 GPa and temperatures of 1750°C and above, whole rock melting takes place (Stöffler et al., 1991).

In addition, the results are also of interest for the upcoming BepiColombo mission to Mercury, for which a comprehensive database of all types of planetary materials will be compiled in the Berlin Emissivity database (BED) (Maturilli et al., 2008; Hiesinger et al., 2010). Since the surface of Mercury is covered by regolith as a result of abundant impacts, a high abundance of impact melts and glassy components can be expected to occur on its surface (Hörz and Cintala, 1997; Benkhoff et al., 2010; Stockstill-Cahill et al., 2014; Spray, 2016). Considering the surface of Mercury the impact rate and velocities are much higher than on the Moon, therefore much more impact melt and glassy components will occur (Cinatla, 1992). Thus, it is important to provide spectral data in order to investigate the dependence of spectral properties from the degree of shock metamorphism and shock-melting of mafic materials (as found within ordinary chondrites).

## 2. Samples and Techniques

### 2.1 Samples

For this study, 9 samples of the Chelyabinsk (Chel) meteorite were obtained from the meteorite collection at the Institut für Planetologie (Westfälische Wilhelms-Universität Münster). Polished thin sections (PL13025 (Chel-1), PL13026 (Chel-2), PL13028 (Chel-4), PL13029 (Chel-5), PL13041 (Chel-6), PL13042 (Chel-7), PL13053 (Chel-12), PL16020 (Chel-16), PL16021 (Chel-17)) were used for characterization through optical microscopy and SEM/EDX, as well for in-situ FTIR studies.

From texturally different Chelyabinsk fragments (Chelabinsk-12, -16, and -17) bulk material was separated into the three main recognizable lithologies: (a) moderately shocked light material without abundant shock veins (e.g., Figs. 1a and 2; Chelyabinsk-12), (b) shock-darkened material (e.g., Figs. 1f and 3; Chelyabinsk-17), and (c) a fraction of impact melt lithology (e.g., Figs. 1e and 4; Chelyabinsk-16) having masses of 1.5 g, 0.8 g, and 1.0 g, respectively. In order to obtain grain size fractions, the sample material was crushed in a steel mortar and an agate mortar. The material was cleaned using acetone to remove potential terrestrial contamination. Following crushing and cleaning, four grain size fractions were sieved from each lithology sample: 0-25 µm, 25-63 µm, 63-125 µm, and 125-250 µm.

Since the various lithologies occur very close to each other in the samples, it was difficult to separate 'pure' material for each lithology for the size fractions. Especially difficult to visually distinguish are the impact melt lithology and the shock-darkened fraction. In order to confirm the bulk composition of the grain size fractions for the shock-darkened and impact melt lithologies, grains of the 125-250 µm fractions were embedded in resin and polished. Microscopic identification confirmed that over 70% of the grains in the shock-darkened lithology are in fact material from this lithology; the remaining portion is made up of fragments from the other two lithologies. The fraction of the impact melt lithology is very homogeneous.

## 2.2 Optical and Petrological Microscopy

Overview images were obtained using a KEYENCE Digital Microscope VHX-500F (Fig. 1). For the first petrological studies, samples were studied by optical microscopy in transmitted and reflected light using a Zeiss Axiophot.

Using this technique, the three main lithologies could well be discriminated and the relationships among each other were documented (Figs. 1,2).

**2.3 Scanning Electron Microscopy and Analysis (SEM/EDX)**

A JEOL 6610-LV electron microscope (SEM) at the Interdisciplinary Center for Electron Microscopy and Microanalysis (ICEM) at the Westfälische Wilhelms-Universität Münster was used to study the different lithologies of Chelyabinsk in detail (Figs. 3-5), in particular the fine-grained texture of the impact melt lithology, in order to identify and chemically characterize its mineral phases. The attached EDS system (INCA; Oxford Instruments) was used for chemical characterization and analysis of the different mineral phases at 20 keV. The beam current was controlled using a Faraday Cup. Standard (Astimex) olivine (Mg, Fe, Si), jadeite (Na), plagioclase (Al), sanidine (K), diopside (Ca), rutile (Ti), chromium-oxide (Cr), rhodonite (Mn), and pentlandite (Ni) were used as natural and synthetic standards.

**2.4 Powder FTIR-Analyses**

In order to ensure diffuse reflectance, powders from each size fraction were placed in aluminum sample cups (1 cm diameter), with their surface flattened using a spatula following a similar procedure described by Mustard and Hayes (1997). We used a Bruker Vertex 70 infrared system with a MCT detector for the bidirectional reflectance analyses in the mid-infrared from 2.5-19 µm. These analyses were made at the IRIS (Infrared and Raman for Interplanetary Spectroscopy) laboratory at the Institut für Planetologie (Münster).

Measurements were made under low pressure ($10^{-3}$ bar). For a high signal-to-noise ratio, 512 scans were accumulated for each size fraction. A commercial diffuse gold standard (INFRAGOLD$^{TM}$) was applied for background calibration. We obtained analyses in a variable geometry stage (Bruker A513) for the MERTIS database. This is necessary to emulate various observational geometries of an orbiter. The results in this study were obtained at 30° incidence (i) and 30° emergence angle (e). To compare them with remote sensing data in the thermal infrared range, reflectance data have to be converted into emission. This is done using Kirchhoff's law: ε = 1 – R (R=Reflectance, ε = Emission) (Nicodemus, 1965). This relationship works well for comparing directional emissivity and directional hemispherical reflectance (Hapke, 1993; Salisbury et al., 1994). However, for a direct comparison of directional emissivity with reflectance by using Kirchhoff's law, the reflected light in all directions has to be collected (Salisbury et al., 1991a; Thomson and Salisbury, 1993; Christensen et al., 2000; King et al., 2004). In this study, a bidirectional, variable mirror set-up was used without a hemisphere integrating all radiation. This has to be considered when the results of this study are used in a quantitative comparison with emission data (Salisbury et al., 1991a; Christensen et al., 2000).

We present powder spectra for all size fractions from 7-18 µm (online appendix, Tab. A1a). Spectra are presented in reflectance, from 0-1 (Fig. 6a-c). For the features below 7 µm, important for volatile species, we show representative spectra of the largest size fraction (125-250 µm) in Fig. 7a-c.

For comparison purposes, we used reference spectra from the NASA RELAB (Reflectance Experiment Laboratory) Database at Brown University (Pieters, 1983; Miliken et al., 2016) and the Johns Hopkins ASTER laboratory (Baldridge et al., 2009).

The results of this study will be made available through the IRIS database in Münster (http://www.uni-muenster.de/Planetology/ifp/ausstattung/iris_spectra_database.html) and the BED (Berlin Emissivity Database) at the DLR in Berlin. This database is compiled for the interpretation of mid-infrared spectra form the MERTIS (Mercury Thermal Radiometer and Thermal Infrared Spectrometer) instrument onboard of the upcoming ESA/JAXA BepiColombo mission to Mercury and for other future space missions.

## 2.5 In-situ FTIR Micro-spectroscopy

Given the heterogeneous character of the lithologies, we also obtained additional in-situ infrared spectra from the lithologies in thin sections. This was of interest in order to get analyses of spots of 'pure' material in contrast to the diffuse reflectance analyses of bulk materials of the three lithologies, which are by nature mixtures of many components. For in-situ analyses of the moderately shocked light material it was relatively easy to find 'pure', homogenous spots (Fig.7). On the other hand, in the micro-FTIR analyses of shock-darkened material and of the impact melt lithology, it was more difficult to avoid fragments from the other lithologies, we tried to obtain spectra from relatively homogeneous spots of 1 mm$^2$ (Fig.7).

We used a Bruker Hyperion 2000 IR microscope attached to the external port of a Bruker Vertex 70v at the Hochschule Emden/Leer for in-situ analyses on the thin sections. An aperture of 1 mm$^2$ was used to perform the analyses and in each case 128 scans were added. Labelled rectangles in Figure 1 show areas studied with micro-FTIR. For background calibration, a gold mirror was used. Results are presented in Fig. 7a-c and Tab. A2a-c.

The diffusely scattered part of the light is very small in micro-FTIR in specular reflectance mode, since highly polished samples were analyzed. Specular reflectance can be related to the directed emission, as required by Kirchhoff's law, so the differences will be mainly in spectral contrast. Band shape and positions of the spectra will be comparable between emission and converted specular reflectance (Ramsey and Fink, 1999; Byrnes et al., 2007; Lee et al., 2010).

## 3. Results

3.1 Optical, Polarization Microscopy and Scanning Electron Microscopy Analysis (SEM/EDX)

Figures 1a-f show optical overviews of the petrologic thin sections of the samples used in this study. The three lithologies can already be quite well distinguished in the optical images. The moderately shocked lithology of basically unaltered LL5-, LL5/6-, or LL6-material is characterized by its light appearance (e.g., in samples Chelyabinsk 2, 4, and 5; Figs. 1a, b, and c, respectively; Fig. 2), and is dominated by olivine (Righter et al., 2015). In contrast, shock-darkened material is nearly black, as shown in the samples of Chelyabinsk 1, 4, 5, 6 and 7 (Figs. 1b, d, c, e, and f; Fig. 3). Finally, the impact melt lithology can be recognized by its greyish appearance typically having abundant small fragments embedded within the impact melt. Typical examples for this lithology are observed within the meteorite pieces of Chelyabinsk 1, 4, and 6 (Figs. 1b, d, and e; Fig. 4). The impact melt lithology exists as large, up to centimeter-sized units (Fig. 1e) as well as in the form of thin two-dimensional shock veins often observed within the light, moderately-shocked lithology (Figs. 2 and 4c).

Polarization microscopy in reflected light and SEM-studies clearly help to distinguish between the shock-darkened rock type and the impact melt lithologies. The melt lithology shows a typical texture with feldspar-normative melt embedding mafic constituents, metal-sulfide intergrowth (spherules), and lithic fragments. The fragments represent the moderately shocked ordinary chondrite materials and were incorporated into the melt (Fig. 4). The occurrence of generally round metal-sulfide intergrowth is typical for impact melts in chondrites, while in the shock-darkened lithologies sulfides and metals occur on the grain boundaries of the appropriate mineral phases or fill up cracks and cleavage planes of the participating minerals (Fig. 3). We interpret the texture with the orientations of the networks in Chelyabinsk-4 that they may have been formed by different impact processes (Fig. 3b). The newly crystallized olivines within the melt lithology are zoned (Figs. 4 and 5). They have cores of $Fa_{13-20}$ with rim compositions similar to that of the olivine of the target rock (~$Fa_{29}$).

## 3.2 Diffuse Reflectance and Micro-FTIR

### 3.2.1 Impact Melt Lithology

The Christiansen feature (CF) is a characteristic reflectance minimum that can be used as a compositional indicator for the bulk composition of a sample. There is a good correlation between the $SiO_2$ content and the position of the CF, which shifts to higher wavelengths with decreasing $SiO_2$ concentration. This allows distinguishing various rock types based on this indicator, for example between acidic and basic rocks (Salisbury, 1993; Cooper et al., 2002). In the impact melt lithology (Chelyabinsk-16; Fig. 6a; Tab. A1a) the CF is found at 8.4 to 8.7 µm.

Bulk powder spectra obtained from the impact melt lithology (Chelyabinsk-16; Fig. 6a; Tab. A1a) can be described as simple spectra dominated by one strong, broad reststrahlen band (RB) feature and is located at 11.2-11.3 µm. Minor RB occur at 9.7-9.8 µm, 10.3 µm, 12 µm, and 16.3-16.4 µm. The transparency feature (TF), characteristic for the smallest grain size fractions (Salisbury, 1993) is at 13.1 µm, and the weak water feature is located at 2.8-2.9 µm (Table A1b). Chelyabinsk is a fall, but the samples were collected some time afterwards and may have been exposed to terrestrial weathering. So the minor volatile bands in most spectra are likely results from terrestrial adsorption, we omitted this spectral range part in the figures of the bulk powder spectra. Band positions are presented in Table A1b.

Micro-FTIR analyses of the impact melt lithology provide a series of spectra (Fig. 7a; Tab. A2c), which show features similar to the spectra obtained from the bulk analyses: a CF is found between 8 and 8.1 µm, a dominating RB at 11.2-11.3 µm, and minor RB at 9.7-9.8 µm, 10.3 µm and 12 µm. At longer wavelengths, only very weak bands are seen. However, the spectra show (slight) differences among each other, mainly based on the occurrence of the feature at 9.7–9.8 µm, which only appears in half of the spectra of this type.

Thus, we identified a sub-group in the melt spectra of samples, with the simplest spectra exhibiting the fewest features as expected from possibly amorphous, glassy material without significant crystalline inclusions. These 'pure' impact melt-like spectra (Fig. 7a; Tab. A2c) only have a shoulder in the 9.6-9.8 µm range. Otherwise the spectral characteristics for this sub-group include a CF at 8-8.1 µm, the dominating RB at 11.2-11.3 µm, and weaker RB at 10.3 µm and 12 µm.

### 3.2.2 Shock-Darkened Lithology

The spectra obtained from the powdered bulk shock-darkened lithology (Chelyabinsk-17; Fig. 6b; Tab. A1a) have similar characteristics to those of the moderately shocked light material. A 'twin-peaks' feature between 10.8 and 11.3 µm is typical, accompanied by two smaller RB features at 9.6-9.7 µm and 10.3 µm. Further RB features are seen at ~16.7 µm. The transparency feature is present at 12.9 µm and the CF is at 8.3-8.8 µm (Fig.6b).

The micro-FTIR spot analyses of the shock-darkened areas (Fig. 7b; Tab.A2b) in thin sections also show spectral variations. In addition to spectra characteristic of the bulk shock-darkened lithology, several are similar to those obtained from the impact melt lithology. The shock-darkened lithology has the CF at 8.1-8.7 µm. Dominating RBs are at 10.7-10.8 µm and 11.2-11.3 µm, as well as a feature with strongly varying intensity at 9.2-9.3 µm (Fig. 7b; Tab.A2b). Furthermore, minor bands occur at 9.6-10 µm, 10.3-10.5 µm, and 12 µm. At longer wavelengths, two significant bands are clearly detectable at 13.8-13.9 µm and 14.5-14.8 µm.

The micro-FTIR spectra of this lithology from spots that are similar to those of impact melt lithology material (Fig. 7b; Tab.A2b) have their CF at 8.1-9.1 µm with the strong RB at 11-11.3 µm. Minor features are visible at 9.5-10 µm, 10.3 µm, and 12 µm. These spectra probably represent shock-produced melts typically found in the shock veins and that are difficult to recognize in the shock-darkened lithology.

**3.2.3 Moderately Shocked, Light Lithology**

The bulk powder sample of the dominating moderately shocked, light lithology (Chelyabinsk-12; Fig. 6c; Tab. A1a) has its strongest reststrahlen bands as a 'twin peaks' feature between 10.8 µm and 11.2-11.3 µm. Further significant RB features occur at 9.1 µm, 9.5, and 10.3 µm. At longer wavelengths, additional RBs are seen from 16.2 µm - 16.9 µm. The CF varies between 8.5-8.8 µm and the TF is at 13 µm.

The micro-FTIR analyses of the moderately shocked, light lithology (Fig. 7c; Tab. A2a) reveal more variations compared with the bulk analyses. Again, we divided the spectra of these spots into two groups. (1) Spectra (less abundant), which are related to those of the impact melt lithology and which are dominated by one strong RB, and (2) those with the 'twin-peak' feature and more observable peaks in general.

In the latter group (2) (Fig. 7c; Tab. A2a), the strongest two reststrahlen bands are at 10.7-10.9 µm and 11.2-11.3 µm. A RB at 9.2 µm varies from being the strongest band (Chel4_3) to a very weak feature (e.g., Chel2_3).

Furthermore, minor features with varying intensity are seen at 9.6-9.8 µm, 10 µm, and 10.2-10.5 µm. Less significant features at longer wavelengths occur at 12 µm, 13.8, and 14.5-14.6 µm. The CF is between 8.2 and 8.7 µm.

The spectra from spots obtained by micro-FTIR that look similar to those obtained from the impact melt lithology (1) (Fig. 7c; Tab. A2a) have a main band at 10.8-11.2 µm accompanied by further RBs at 9.6-9.7 µm, 10.3-10.4 µm, and 12 µm. There are various significantly weaker features at wavelengths greater than 12 µm. The CF is found between 8.6 and 9.8 µm. These spectra probably represent shock-produced melts found in the shock veins that cross cut the moderately shocked, light lithologies (Figs. 1b and 2).

## 4. Discussion

### 4.1 Mineralogical characteristics and peculiarities of Chelyabinsk

Chelyabinsk is not only another ordinary LL5 chondrite, but a spectacular chondrite breccia containing light-colored fragments of LL5-lithology as well as even stronger metamorphosed fragments (LL6). These light-colored fragments can have various abundances of thin shock veins (Fig.1). Besides these light clasts, a shock-darkened type of fragment exists as well as an impact melt lithology. Chelyabinsk is a well-lithified breccia, in contrast to Almahata Sitta

(e.g., Bischoff et al., 2010, Goodrich et al., 2014; Horstmann and Bischoff, 2014), which contains fragments that were loosely-packed in the asteroid prior to the entry and break-up in Earth's atmosphere. Based on the brecciated nature of the sample and the occurrence of fragments with a different degree of metamorphism, the Chelyabinsk meteorite is better classified as a LL5-6 chondrite breccia instead of a LL5 chondrite (Bischoff et al., 2006, 2013).

### 4.2. Spectral characteristics of different lithologies

Considering the RB features observed in the powder spectra obtained from the moderately shocked, light lithology (Fig. 6c; 8), the bands at 9.5 µm, 10.3 µm, 10.8 µm, and 11.2-11.3 µm, and minor features at 12 µm (Tab.A1a) are consistent with fayalite (Fa) content (27-29 mol%) of olivine for the light lithology (Hamilton, 2010; Lane et al., 2011; Kohout et al., 2014; Ruzicka et al., 2015; Righter et al., 2015). This result holds for both the bulk and thin section spectra (Tab.A2c).

Similarly, the pyroxene-related RB in the powdered bulk at 9.1 µm, 9.5 µm, and 10.3 µm (Fig.6c; 8), as well as the micro-FTIR analyses generally corroborate the (ferrosilite) $Fs_{22-23}$, composition of low-Ca pyroxene (Hamilton, 2000; Kohout et al., 2014; Righter et al., 2015; Ruzicka et al., 2015). However, in several cases they overlap with the olivine features (Tab.A1a; 2c).

In the bulk powdered shock-darkened lithology (Fig. 6b; 8), the situation is similar: features at 9.6-9.7 µm, 10.3 µm, 10.7-10.8 µm and 11.3 µm (Tab.A1a) are overlapping olivine and pyroxene RB (Hamilton, 2000; 2010; Lane et al., 2011) confirming similar forsterite (Fo) and Fs contents compared with those in the moderately shocked, light lithology (Kohout et al., 2014; Righter et al., 2015). The situation for the micro-FTIR analyses is similar (Tab. A2b).

The broad dominating RB observed in spectra from the impact melt lithology (Fig. 6a; 8) covers the range of the twin peaks features in the other lithologies from 10.8-11.3 µm. The peak position at 11.2-11.3 µm (Tab. A1a) is characteristic for olivine (Hamilton, 2000; 2010; Lane et al., 2011). Olivine features occur at 9.7 µm and 12 µm (Tab. A1a), and overlapping olivine and pyroxene features at 10.3 µm (Fig. 6a; 8)(Hamilton, 2000; 2010). This confirms the SEM observations of small olivine crystals embedded in the probably amorphous, plagioclase-normative groundmass (Fig. 4). For the micro-FTIR analyses, the situation is similar (Tab. A2c). The spectrum is clearly different from bands obtained from basaltic glasses, which fall in the 10-10.4 µm region (Dufresne et al., 2009; Minitti and Hamilton, 2010). This indicates that we see a mixture of spectral features from several phases.

The CF observed in spectra of the finest size fractions (Fig. 9; Tab. A1a) obtained from the three lithologies (8.7-8.8 µm) - in correlation with the position of the TF (12.9-13.1 µm) - point toward olivine as the most common phase (Salisbury, 1993). The range of the CF in spectra of the larger grain size fractions (>25 µm) of the bulk samples (8.3-8.6 µm) confirms this finding (Fig.8) (Salisbury, 1993).

The CFs in the micro-FTIR analyses (Fig. 7a-c; Tab.A2a-c) show a wider range than the bulk fractions, and are often recorded at significantly shorter wavelengths. A few CFs, especially from spots analyzed within the melt lithology, fall in the spectral region that is typical of feldspar-rich components (<8.2 µm) (Salisbury, 1993)(Fig. 7a-c; Tab. A2a-c), indicating a mixture with abundant plagioclase. This points toward a high abundance of the feldspathic groundmass within the shock-melted lithologies (Fig .4).

The occurrence of Fo-rich olivine (Fa$_{13-20}$; Fig. 5) within the cores of the newly-crystallized olivine does not show up in the obtained spectra. The abundance of olivine with this composition is too low (20 vol% estimated) and the chemical difference to the Fa$_{29}$ olivine of the bulk rock is probably too small to yield a peak-position shift in the spectra.

One potential limitation in using Micro-FTIR are orientation effects of the crystals affecting the spectral features (Benedix et al., 2015 and 2016; Morlok et al., 2006). In the case of the powder spectra we can expect randomly oriented grains, which average out orientation effects. In the case of the Micro-

FTIR analyses on thin sections, the aperture used was 1 mm$^2$. Grain sizes in the various lithologies analyzed are all well below this size. Especially the impact melt lithology – excluding the incorporated fragments from the target rock - is very fine-grained with the silicates usually in the 10-30 µm size range (Fig.4). This characteristic probably averages out orientation effects as indicated by the similarity of all micro-FTIR spectra obtained from the impact melt lithology, which are very similar to each other (Fig.7a). Both the shock-darkened and the moderately shocked light lithology exhibit larger grain sizes (Fig.2,3). Thus, the larger variation seen in the micro-FTIR spectra could therefore be somewhat related to orientation effects (but which again are largely evened out in the bulk spectra). However, a series of spectra in both the shock-darkened and moderately shocked light lithologies exhibits the very similar features of the impact melt lithology (Fig.7b,c), confirming the homogeneity at least of this lithology even on very small scales.

**4.3. Dependence of the degree of impact shock and spectral features**

In order to identify and characterize the effects of shock melting and darkening, the spectra of the 125-250 µm grain size separates of the three lithologies are compared (Fig. 8). For this purpose, we focus mainly on the powdered bulk material, since it is more representative of the whole lithology than various individual spots and thus more applicable to remote sensing applications. We picked the spectrum of spot Chel6_1 (Fig. 7a) obtained via micro-FTIR as an example of a very 'pure' impact melt lithology. However, it has to be kept in mind that the micro-FTIR data is from polished samples, resulting in much higher intensities compared to the powder analyses.

The spectrum of the bulk moderately shocked, light lithology is characterized by five clear features in the 8-14 µm range (Fig.8). The light-colored clasts of metamorphosed type LL5 or LL6 lithologies have experienced shock at levels near S4 (Stöffler et al., 1991). This indicates an equilibrium shock pressure of at least 15-20 GPa and a post-shock temperature increase of 100-150 °C. During this shock pressure the formation of shock veins and melt pockets is often observed indicating the effects of local pT-excursions (Stöffler et al., 1991), which explains the micro-FTIR spectra similar to the impact melt lithology (Fig.7a-c). Thus, the spectral features of the bulk moderately shocked, light lithology were probably not significantly changed by the impact shock in the 15-20 GPa region, and representative of S4:The shock-darkened lithology may have formed under somewhat more intensive shock pressures compared with the moderately shocked, light lithology. The darkening is caused by the formation of metal-troilite veins along grain boundaries and fractures of olivine and pyroxene (Fig. 3). Here we already observe first changes in the spectral features (Figs. 6b,7b,8): The shock-darkened bulk material shows two spectroscopic features at 9.1-9.2 µm and 9.6-9.7 µm, basically turning into shoulders, while the characteristic 'twin peak' at 10.8 – 11.3 µm remains (Tab.A1a). This could be indicative for the subsequent shock stage S5.

The bulk spectrum of the impact melt lithology has the twin-peaks feature forming one big band at 11.2 µm (Figs. 6a,7a,8; Tab.A1a). The loss of this feature could be the result of peak broadening due to amorphization and weak crystallinity from a significant increase of amorphous material (Johnson et al., 2007; Hamilton et al., 2010; Lane et al., 2011; Jaret et al., 2015).

However, the remaining bands at ~9.7µm, 10.3 µm, 12 µm, and ~16 µm (Fig.8; Tab.A1a) show that there are indeed still abundant crystalline species - 'contamination' of the pure impact melt material by fragments of the two other lithologies, or material crystallized from the melt. Since the incorporation of small fragments of light materials are unavoidable in bulk analyses, the best way to find 'pure' impact melt material is to look at the micro-FTIR spectra. Spot Chel6_1 (Fig. 7a) shows only a few features: In addition to the loss of the twin-peak band, it is also characterized by the loss of the feature at 9.6-9.8 µm and the shoulder at 9.1 µm. There are still strong crystalline features visible at 10.3 µm and 12 µm (Tab.A2c), probably as a result of the abundant small olivine that crystallized in the impact melt lithology (see Figs. 1e, 4, and 5).

However, the 9.7-9.8 µm olivine feature is also sensitive to the specific analytical technique. While the feature occurs throughout most of the forsterite-fayalite series, it fades in specular reflectance analyses of polished material into a shoulder between $Fa_{30}$-$Fa_{25}$ (Hamilton, 2010; Lane et al., 2011). This is slightly below the composition of the olivine in the melt veins (Fig. 5), but in the range of the unmelted material. Thus, the loss of the 9.7-9.8 µm feature alone is not indicative of an amorphization of the material. On the other hand, various other micro-FTIR spectra, not just of the pure melt areas, show actual features in this position (Fig. 7a-c; Tab. A2a-c), which are probably not pyroxene features (Hamilton, 2000). This would indicate that we do not see an instrumental artifact that explains the fading of a feature in this region as observed by Lane et al. (2010). Still, the loss of the twin-peak band at 10.8-11.3µm is probably the most diagnostic sign of the formation of amorphous material, and thus is the most useful as spectral indicator for an impact melt lithology.

The formation of the impact melt lithology required higher shock pressures. Stöffler et al. (1991) estimated an equilibrium peak shock pressure for whole rock melting on the order of 75-90 GPa (S6) with a minimum temperature increase of 1500 °C. Righter et al. (2015) also estimated that the impact melt breccia lithology formed at high temperatures of ~1600 °C and that it experienced rapid cooling. Therefore, the loss of the 'twin-peak' and the 9.7-9.8 µm feature could be indicator for at least a shock stage of S6 (Fig.8).

Mid-infrared studies of experimentally shocked minerals and rocks show varying sensitivity for changes with increasing shock pressure. Pyroxenite exhibits only minor changes even at pressures up to 63 GPa (Johnson et al., 2002). Basalts are more susceptible to changes due to their feldspar abundances, which show many spectral changes during shock metamorphism. Feldspars exhibit loss of features at pressures up to 50 GPa (Stöffler and Hornemann, 1972; Ostertag, 1983; Johnson et al., 2003, 2007, 2013; Jaret et al., 2015; Martin et al., 2016). Thus, the loss of the features in the ordinary chondrite spectrum takes place in a different pressure range than in the experimentally shocked materials, probably due to the differences in petrology (high olivine and pyroxene abundance) of the ordinary chondrites.

### 4.4. Comparison with other LL Chondrites

A comparison of the finest grain size fraction with earlier studies of metamorphosed LL chondrites shows basically identical spectra (Fig. 9). The spectrum of the size fraction <125 µm of the Tuxtuac LL5 meteorite has the CF at 8.6 µm, the dominant RB at 10.8 µm and 11.2 µm, and further RBs at 9.5 µm and 10.3 µm. The TF is rather broad (due to the larger grain size range) with two bands at 12.3 and 12.9 µm. This is very similar to the analyses for the moderately shocked light and shock-darkened Chelyabinsk lithologies (Tab.A1a).

On the other hand, the spectrum obtained from the grain size fraction <250 µm of the LL6 chondrite Paragould (Fig.9) shows no TF. The CF is at 8.6 µm, the two strong RBs are at 10.8 µm and 11.3 µm, and a further strong band is at 10.3 µm. The feature at ~9.6 µm is only weakly resolved, indicating a higher abundance of shock-darkened material (Tab.A1a).

4.5 Comparison bulk powder and micro-FTIR analyses

Due to the variety of features the quantitative comparison between micro-FTIR and bulk powder spectra of the moderately shocked light lithology and the shock-darkened lithology is difficult. Therefore, we focus on the homogeneous impact melt lithology for this comparison.

The intensity ratios normalized to the dominating 11.2-11.3 µm feature and the minor features vary between the powder and micro-FTIR analyses. The feature at 9.7 µm has an intensity of 64% to 74% of the 11.3 µm band in the powder spectra, and 26% to 42% in the micro-FTIR data (Tab. A1a, A2c). The ratio for the 10.3 µm band has 73% - 75% for powder and 43% - 61% for micro-

FTIR). The band at 12 µm has a different behavior, here intensities are 52% - 53% for powder spectra and 54% - 60% for micro-FTIR (Tab. A1a, A2c).

This observation points towards a relative increase of the intensity of these minor features with increasing wavelength in micro-FTIR compared to powder analyses.

In the wavelength range covered by both techniques, only the CF shows a difference – 8.0-8.1 µm for the micro-FTIR, but 8.4-8.7 µm in the bulk powder measurements (Tab. A1a, A2c). This difference could be explained with small, compositionally different fragments in the melt, but also result of the different techniques used, bulk and micro-FTIR (compare Cooper et al., 2002). However, the other bands are at basically identical positions in both techniques. For the other two lithologies the range of the CF for both techniques overlap (Tab. A1, A2).

This mainly confirms the finding of earlier studies comparing analyses of powders and polished material (Klima and Pieters, 2006; Benedix and Hamilton, 2007).

## 5. Summary & Conclusions

We have provided mid-infrared spectra of three typical, well characterized lithologies from the Chelyabinsk ordinary chondrite meteorite. With increasing influence of impact shock from 'pristine' LL5 or LL6 material with low or moderate degree of shock to shock-darkened and impact melt lithologies as the endmember, we observe the fading/disappearing of spectral features.

Most prominent is the loss of a 'twin peak' feature between 10.8 and 11.3 µm, which turns into a single peak feature. In in-situ analyses of very 'pure' areas in the impact melt lithology, the features at ~9.6 µm and ~9.1 µm are also lost. Glass still retains short range order in form of the $SiO_4$ tetrahedra, which are randomly linked due to the loss of lattice structure and thus long-range order. While the well-defined bond angles and lengths in crystalline materials show narrow infrared bands, the random distribution in glass causes band broadening in glass as result of many similar vibrational modes (e.g. Farmer, 1974; Speck et al., 2011). The broad feature observed in silicate glasses in the region around 10 µm is the result of various asymmetric stretching Si-O stretching modes (Dalby and King, 2006; Dufresne et al., 2009). Thus, these losses of features are probably correlated with decreasing amounts of crystal structure with increasing shock and melting.

This may allow to connect mid-infrared features with the range of stages for shock metamorphism based on changes in the petrology (Stöffler et al.,1991). Changes up to shock stage S4 would be minor, the shock darkened lithology could represent S5 and the impact melt lithology S6 and higher. These findings will be useful in remote sensing studies, since they could allow the identification of impact melts on parent bodies of ordinary chondrite material. This could also help determining degrees of shock on an asteroidal surface. However, asteroidal regoliths can be expected to consist of intimate mixtures of rocks that underwent the whole range of impact shock metamorphism (Hörz and Cintala, 1997). So 'bulk' spectra cannot be expected to only show features of one group of shocked materials. Such studies of the degree of impact metamorphism have to take all lithologies into account using quantitative deconvolution algorithms to calculate the abundance of the different lithologies from remote sensing data based on their infrared spectra (e.g. Ramsey et al., 1998, Rommel et al., 2014).

Alternatively, high-resolution mapping of an asteroidal surface comparable to e.g. the OSIRIS-REx Thermal Emission Spectrometer (OTES) (Lauretta et al., 2012) might allow identifying spots on the surface dominated by spectral features of lithologies such as impact melt glass.

Finally, micro-FTIR allowed us to separate the spectral features of the various phases that are often mixed in the bulk spectra. Even the bulk material of the impact melt lithology still contains small fragments of other materials (Fig. 4b) that are difficult to separate manually. With the help of the in-situ analyses, it was possible to analyze pure 'melt' products, thus providing valuable additional spectral information difficult to obtain with bulk techniques. With the recent popularity of this micro-analytical technique (Glotch et al., 2011; Hamilton and Connolly jr., 2012; Jaret et al., 2015; Benedix et al., 2016; Hamilton et al., 2016; Martin et al., 2016), such data will prove valuable. However, it has to be taken into account for quantitative comparisons that relative band intensities of bulk powder analyses and those of micro-FTIR differ, while band positions remain very similar.

The Chelyabinsk spectra are very similar to the mid-infrared spectra of other LL chondrites. This indicates that the findings of this study can be probably be applied to other LL chondrites.

## 6. Acknowledgements

We thank Isabelle Dittmar (Emden) for analytical support, Ulla Heitmann (Münster) for thin section preparation, and Celeste Brennecka (Münster) for editorial support. We also thanks the two reviewers, G. Benedix and V. Reddy and editor W.M. Grundy for helping improving this manuscript.

This work was partly supported by DLR grant 50 QW 1302 in the framework of the BepiColombo mission, and the German Research Foundation within the Priority Program "The First 10 Million Years – a Planetary Materials Approach" (SPP 1385). This research utilizes spectra acquired by C.M. Pieters and J.M. Sunshine with the NASA RELAB facility at Brown University.

**Tables are moved into the Appendix**

| Moderately shocked light lithology | CF | | | | | | | | | | | | | TF | | | | | | | | | | |
|---|---|---|---|---|---|---|---|---|---|---|---|---|---|---|---|---|---|---|---|---|---|---|---|---|
| 0-25 µm | 8,80 | 9,10 | **61** | 9,53 | **78** | | | 10,27 | **76** | 10,77 | **99** | 11,25 | **100** | | | 12,99 | **98** | 13,99 | **45** | 16,20 | **45** | 16,54 | **45** | 16,73 | **45** |
| 25-63 µm | 8,64 | | | 9,54 | **70** | | | 10,28 | **90** | 10,76 | **100** | 11,23 | **94** | 11,97 | **42** | | | | | | | | | 16,76 | **34** |
| 63-125 µm | 8,57 | | | 9,54 | **68** | | | 10,28 | **78** | 10,76 | **100** | 11,24 | **94** | 11,97 | **43** | | | | | | | | | 16,66 | **39** |
| 125-250 µm | 8,49 | | | 9,54 | **72** | | | 10,28 | **80** | 10,77 | **100** | 11,25 | **96** | 11,97 | **48** | | | | | | | | | 16,90 | **40** |
| Shock darkened lithology | | | | | | | | | | | | | | | | | | | | | | | | | |
| 0-25 µm | 8,77 | | | 9,56 | **72** | | | 10,29 | **75** | 10,84 | **95** | 11,27 | **100** | | | 12,87 | **94** | 13,95 | **73** | 16,05 | **51** | 16,27 | **51** | 17,42 | **49** |
| 25-63 µm | 8,50 | | | 9,59 | **63** | 9,94 | **61** | 10,29 | **77** | 10,81 | **98** | 11,26 | **100** | 11,98 | **53** | | | | | | | | | 16,66 | **38** |
| 63-125 µm | 8,47 | | | 9,65 | **61** | | | 10,29 | **76** | 10,81 | **98** | 11,26 | **100** | 11,97 | **52** | | | | | | | 16,58 | **36** | 16,73 | **36** |
| 125-250 µm | 8,3 | | | 9,61 | **64** | | | 10,29 | **79** | 10,79 | **98** | 11,27 | **100** | 11,98 | **56** | | | 13,79 | **31** | | | | | 16,75 | **40** |
| Impact melt lithology | | | | | | | | | | | | | | | | | | | | | | | | | |
| 0-25 µm | 8,7 | | | 9,71 | **74** | | | 10,25 | **74** | | | 11,26 | **100** | | | 13,05 | **91** | | | 15,99 | **57** | | | 17,29 | **50** |
| 25-63 µm | 8,48 | | | 9,74 | **65** | | | 10,27 | **75** | | | 11,23 | **100** | 11,98 | **53** | | | | | | | 16,32 | **36** | | |
| 63-125 µm | 8,51 | | | 9,75 | **64** | | | 10,28 | **75** | | | 11,23 | **100** | 11,98 | **52** | | | | | | | 16,34 | **33** | | |
| 125-250 µm | 8,38 | | | 9,74 | **64** | | | 10,27 | **73** | | | 11,23 | **100** | 11,98 | **53** | | | | | | | 16,39 | **33** | | |

Table A1a: Band position bulk powder FTIR studies. Band positions in µm. Bold number right of RB: Intensity of feature in percent (%) normalized of strongest band

| Moderately shocked light lithology | | | | | | |
|---|---|---|---|---|---|---|
| 0-25 µm | 2,97 | 3,42 | 3,50 | 5,66 | 6,02 | 8,80 |
| 25-63 µm | 2,95 | 3,41 | 3,50 | 5,66 | 6,02 | 8,64 |
| 63-125 µm | 2,97 | 3,41 | 3,50 | 5,65 | 6,02 | 8,57 |
| 125-250 µm | 2,96 | 3,41 | 3,50 | 5,66 | 6,02 | 8,49 |
| | | | | | | |
| Shock darkened lithology | | | | | | |
| 0-25 µm | | | | 5,66 | 6,02 | 8,77 |
| 25-63 µm | | | | 5,66 | 6,02 | 8,50 |
| 63-125 µm | | | | 5,66 | 6,02 | 8,47 |
| 125-250 µm | | | | 5,66 | 6,02 | 8,3 |
| | | | | | | |
| Impact melt lithology | | | | | | |
| 0-25 µm | 2,9 | 3,42 | 3,5 | 5,65 | 6,02 | 8,7 |
| 25-63 µm | 2,8 | 3,42 | 3,5 | 5,65 | 6,02 | 8,48 |
| 63-125 µm | 2,8 | 3,41 | 3,5 | 5,65 | 6,02 | 8,51 |
| 125-250 µm | 2,8 | 3,4 | 3,49 | 5,65 | 6,02 | 8,38 |

Table A1b: Water and volatile features in the bulk powder spectra. Band position in µm.

| Moderately shocked, light lithology | CF | | | | | | | | | | | | | | | | | | | | | | | | |
|---|---|---|---|---|---|---|---|---|---|---|---|---|---|---|---|---|---|---|---|---|---|---|---|---|---|
| Chel2_1 | 8,55 | | | 9,57 | 59 | | | 10,31 | 87 | | | 10,77 | **100** | 11,31 | **97** | 11,95 | **52** | | | | | 13,79 | **21** | 14,57 | **17** |
| Chel2_2 | 8,5 | 9,23 | **79** | 9,57 | **69** | | | 10,24 | **77** | 10,5 | **69** | 10,78 | **77** | 11,3 | **100** | | | | | 13,48 | **17** | 13,81 | **25** | 14,58 | **21** |
| Chel2_3 | 8,6 | 9,17 | **17** | | | 9,97 | **72** | 10,33 | **87** | | | 10,69 | **100** | 11,26 | **98** | 11,97 | **46** | | | | | 13,79 | **19** | 14,53 | **12** |
| Chel2_4 | 8,55 | | | 9,58 | **65** | 9,96 | **44** | 10,33 | **84** | | | 10,74 | **100** | 11,29 | **87** | 11,99 | **47** | 12,6 | **30** | | | | | 14,55 | |
| Chel4_1 | 8,19 | | | 9,76 | **45** | | | 10,33 | **68** | | | 10,87 | **100** | 11,15 | **100** | 11,98 | **50** | | | | | | | | |
| Chel4_3 | 8,5 | 9,16 | **100** | | | 9,98 | **41** | 10,4 | **61** | | | 10,81 | **59** | 11,26 | **61** | | | | | 13,49 | **17** | 13,81 | **21** | | |
| Chel4_4 | 8,17 | 9,22 | **99** | 9,56 | **94** | | | 10,28 | **98** | 10,45 | **96** | 10,77 | **95** | 11,31 | **100** | | | 12,33 | **44** | 12,99 | **33** | 13,48 | **22** | 13,81 | **27** | 14,59 | **24** |
| Chel5_3 | 8,69 | 9,17 | **27** | 9,58 | **36** | | | 10,33 | **57** | | | 10,85 | **95** | 11,24 | **100** | 11,99 | **53** | | | | | 13,8 | **22** | 14,5 | **16** |
| Impact melt-like | | | | | | | | | | | | | | | | | | | | | | | | | |
| Chel4_2 | 8,73 | | | 9,69 | **53** | | | 10,33 | **56** | | | 10,75 | **100** | | | 12 | **41** | | | | | | | | |
| Chel5_1 | 8,7 | | | 9,6 | **37** | | | 10,33 | **53** | | | 10,86 | **100** | | | 11,99 | **45** | | | | | 13,13 | **22** | 13,56 | **17** |
| Chel5_2 | 8,56 | 9,21 | **10** | 9,58 | **31** | | | 10,36 | **46** | | | 10,85 | **100** | | | 11,99 | **39** | 12,51 | **24** | 12,97 | **19** | 13,39 | **15** | 13,81 | **13** |
| Chel2_5 | 9,8 | | | | | | | 10,3 | **39** | | | | | 11,15 | **100** | 11,98 | **39** | | | | | | | | |

Table A2a.

| Shock darkened lithology | CF | | | | | | | | | | | | | | | | | | | | | |
|---|---|---|---|---|---|---|---|---|---|---|---|---|---|---|---|---|---|---|---|---|---|---|
| Chel1_5 | 8,1 | 9,2 | **81** | 9,59 | **82** | 9,74 | **82** | 10,33 | **97** | 10,79 | **100** | 11,25 | **98** | 11,97 | **61** | | | 13,86 | **35** | | |
| Chel1_6 | 8,7 | | | | | | | 10,28 | **86** | 10,77 | **96** | 11,27 | **100** | 11,96 | **56** | | | 13,81 | **27** | 14,58 | **24** |
| Chel6_7 | 8,22 | 9,26 | **63** | | | | | 10,33 | **83** | 10,84 | **96** | 11,29 | **100** | | | | | 13,79 | **30** | 14,8 | **25** |
| Chel6_8 | 8,71 | | | 9,69 | **52** | 9,91 | **54** | 10,31 | **85** | 10,69 | **99** | 11,3 | **100** | 11,99 | **58** | | | | | 14,62 | **18** |
| Chel7_1 | 8,64 | | | 9,64 | **38** | | | 10,32 | **66** | 10,82 | **100** | 11,24 | **98** | 11,98 | **55** | | | 13,75 | **23** | | |
| Chel7_3 | 8,16 | 9,15 | **94** | | | 9,98 | **61** | 10,45 | **90** | 10,79 | **95** | 11,28 | **100** | | | 13,48 | **29** | 13,81 | **32** | 14,54 | **23** |
| Impact melt-like | | | | | | | | | | | | | | | | | | | | | |
| Chel1_3 | 8,27 | | 9,85 | | | | 10,3 | **80** | | | 11,23 | **100** | 11,99 | **58** | | | | | | | |
| Chel1_4 | 8,34 | | 9,68 | | | | 10,31 | **60** | | | 11,22 | **100** | 11,98 | **60** | | | | | | | |
| Chel4_5 | 8,71 | | 9,95 | | | | 10,32 | **63** | | | 11,24 | **100** | 11,97 | **56** | | | | | | | |
| Chel4_6 | 8,12 | | 9,79 | | | | 10,31 | **84** | | | 11,17 | **100** | 11,99 | **61** | | | | | | | |
| Chel5_4 | 8,14 | | 9,65 | | | | 10,32 | **62** | 10,86 | **100** | | | 11,99 | **49** | | | | | | | |
| Chel5_5 | 8,66 | | 9,62 | | | | 10,32 | **71** | 10,63 | **81** | 11,3 | **100** | 11,99 | **62** | | | 13,66 | **27** | | | |
| Chel7_2 | 8,13 | | 9,5 | | | | 10,32 | **53** | | | 11,24 | **100** | 11,98 | **53** | | | | | | | |
| Chel6_6 | 9,13 | | 9,66 | | | | 10,32 | **48** | | | 11 | **100** | 11,96 | **41** | | | | | | | |

Table A2b.

| Impact melt lithology | CF | | | | | | | | | | | | | | |
|---|---|---|---|---|---|---|---|---|---|---|---|---|---|---|---|
| Chel1_1 | 8,06 | 9,78 | **35** | 10,31 | **55** | 11,24 | **100** | 11,98 | **56** | | | | | | |
| Chel1_2 | 8,06 | 9,8 | **39** | 10,31 | **59** | 11,23 | **100** | 11,98 | **59** | | | | | | |
| Chel6_2 | 8,08 | 9,68 | **39** | 10,31 | **61** | 11,2 | **100** | 11,98 | **58** | | | | | | |
| Chel6_3 | 8,06 | 9,74 | **42** | 10,31 | **60** | 11,22 | **100** | 11,98 | **58** | | | | | | |
| Chel6_5 | 8,01 | 9,76 | **35** | 10,3 | **59** | 11,26 | **100** | 11,99 | **60** | | | | | | |
| 'Pure' Impact melt lithology | | | | | | | | | | | | | | | |
| Chel4_7 | 8,03 | 9,78 | **33** | 10,3 | **50** | 11,26 | **100** | 11,98 | **57** | | | | | | |
| Chel4_8 | 8,04 | 9,75 | **37** | 10,3 | **53** | 11,28 | **100** | 11,98 | **57** | | | 13,87 | **23** | 14,86 | **14** |
| Chel4_9 | 8,02 | 9,77 | **27** | 10,3 | **43** | 11,3 | **100** | 11,98 | **54** | 13,13 | **28** | | | | |
| Chel4_10 | 8,03 | 9,77 | **35** | 10,3 | **53** | 11,29 | **100** | 11,98 | **57** | | | | | | |
| Chel6_1 | 8,08 | | | 10,3 | **54** | 11,29 | **100** | 11,98 | **57** | | | | | | |
| Chel6_4 | 8,1 | 9,77 | **26** | 10,3 | **49** | 11,24 | **100** | 11,98 | **55** | | | | | | |

Table A2c

Table A2a-c: Band positions from micro-FTIR studies. Band positions in µm. Bold number right of RB: Intensity of feature in percent (%) normalized to the strongest band.

**Figure Captions**

Fig. 1: Optical images of the polished thin sections analyzed in this study. ML = melt lithology, SD = shock-darkened lithology, F = fragments within the impact melt lithology. Labelled rectangles show areas studied with micro-FTIR.

Fig. 2: Texture and mineralogy of the moderately shocked light lithology in Chelyabinsk; details of Chelyabinsk-2 (a and b; also shown in Fig. 1b) and Chelyabinsk-12 (c, d). These Chelyabinsk fragments show a strongly recrystallized texture (LL5/6 or LL6). Both fragments contain a small number of shock veins (opaque veins) cutting through the rocks. Images in polarized light (crossed Nicols in a-c).

Fig. 3: Textural appearances of the shock-darkened lithology. (a) Chelyabinsk-6: Cracks and cleavage planes of pyroxene ($\sim Fs_{23}$) and olivine ($\sim Fa_{29}$) are filled by troilite (troi) and kamacite (kam). This is not the case for plagioclase ($\sim An_{13}$). (b) Chelyabinsk-4: Within the shock-darkened area of this fragment, very complex networks of metal (kam) and/or troilite (troi) are clearly visible. We interpret the texture with the orientations of the networks that they were formed by different impact processes. (c) Chelyabinsk-17: Grains with the shock-darkened

characteristic in the 125-250 µm grain size separate. (d) Detailed image from the interior of a grain shown in (c). Mainly troilites (troi) and occasionally kamacite (kam) filling fractures in olivine (ol). Images in back-scattered electrons.

Fig. 4: Back-scattered electron (BSE) images of the textures of impact melt lithologies in Chelyabinsk-6 (a,b), Chelyabinsk-4 (c,d), and Chelyabinsk-16 (e,f). (a) Within the impact melt lithology, olivine (~Fa$_{29}$) and pyroxene (~Fs$_{23}$) clasts of the target rocks are embedded within a fine-grained, interstitial matrix, which consists of newly-crystallized olivine (Fa$_{16-27}$). Also enclosed are chromite and metal/troilite assemblages (kam/troi); (b) Detailed image showing an olivine fragment that shows zoning at the boundaries. Fa$_{22-24}$ crystallized onto the interior of the ~Fa$_{29}$-fragment followed by ~Fa$_{28}$ at the outermost edge to the feldspar-normative groundmass (dark grey); (c) Similar texture in the melt matrix of Chelyabinsk-4 as described in Fig. 4a,b, but with more pronounced zoning; (d) A shock vein in Chelyabinsk-4 characterized by the occurrence of abundant shock-produced metal/troilite spherules (troi/kam) embedded within a fine-grained matrix. (e) Grains of the impact melt lithology in the 125-250 µm grain size separate of Chelyabinsk-16. (f) Detailed image from the interior of a grain shown in (e). Olivines (ol) are embedded within a plagioclase-normative mesostasis.

Fig. 5: Olivine within the impact melt lithology. The newly formed matrix olivines are zoned with a Fa-poor core ($Fa_{13-20}$) and a more Fe-rich rim that is similar in composition to that of olivine from the target rocks (~$Fa_{29}$).

Fig. 6a-c: Bulk powder mid-IR (absolute reflectance) analyses of the four grain size fractions from the three lithologies. CF=Christiansen feature, RB=reststrahlen bands, TF=transparency feature.

Fig. 7: Micro-FTIR in-situ analyses of 1 mm$^2$ spots on the thin sections (compare Fig.1a-f). Thin black lines: actual lithologies. Thick, bold lines on top are the 125-250 µm bulk powder analyses (Fig.6a-c) for the three lithologies for comparison.

a) In-situ impact melt lithology. Grey spectra at the bottom are those showing the 'simplest' spectra with fewest features, probably representing the purest impact melt lithology with low abundance of inclusions. b) and c) shock-darkened lithology and moderately shocked light lithology, respectively. Thin grey lines: selected spectra that show similarity with the impact melt lithology.

Fig. 8: Comparison of 125-250 µm grain size fractions of bulk spectra and pure melt vein spectra from micro-FTIR analyses to show changes in relative band intensities with increasing shock. Features between 9 and 10 µm disappear, as does the 'twin peak' feature of the dominating RB near 11 µm. Vertical lines mark the characteristic features.

Fig. 9: Comparison of the finest grain size fractions (<25 µm) of the three Chelyabinsk lithologies with bulk spectra of other LL5 chondrites: Tuxtuac (<125 µm), and Paragould (<250 µm). Vertical lines mark the characteristic features, including the transparency feature.

| Moderately shocked light lithology | CF | | | | | | | | | | | | | | | TF | | | | | | | | | |
|---|---|---|---|---|---|---|---|---|---|---|---|---|---|---|---|---|---|---|---|---|---|---|---|---|---|
| 0-25 µm | 8.80 | 9.10 | **61** | 9.53 | **78** | | | 10.27 | **76** | 10.77 | **99** | 11.25 | **100** | | | 12.99 | **98** | 13.99 | **45** | 16.20 | **45** | 16.54 | **45** | 16.73 | **45** |
| 25-63 µm | 8.64 | | | 9.54 | **70** | | | 10.28 | **90** | 10.76 | **100** | 11.23 | **94** | 11.97 | **42** | | | | | | | | | 16.76 | **34** |
| 63-125 µm | 8.57 | | | 9.54 | **68** | | | 10.28 | **78** | 10.76 | **100** | 11.24 | **94** | 11.97 | **43** | | | | | | | | | 16.66 | **39** |
| 125-250 µm | 8.49 | | | 9.54 | **72** | | | 10.28 | **80** | 10.77 | **100** | 11.25 | **96** | 11.97 | **48** | | | | | | | | | 16.90 | **40** |
| Shock darkened lithology | | | | | | | | | | | | | | | | | | | | | | | | | |
| 0-25 µm | 8.77 | | | 9.56 | **72** | | | 10.29 | **75** | 10.84 | **95** | 11.27 | **100** | | | 12.87 | **94** | 13.95 | **73** | 16.05 | **51** | 16.27 | **51** | 17.42 | **49** |
| 25-63 µm | 8.50 | | | 9.59 | **63** | 9.94 | **61** | 10.29 | **77** | 10.81 | **98** | 11.26 | **100** | 11.98 | **53** | | | | | | | | | 16.66 | **38** |
| 63-125 µm | 8.47 | | | 9.65 | **61** | | | 10.29 | **76** | 10.81 | **98** | 11.26 | **100** | 11.97 | **52** | | | | | 16.58 | **36** | 16.73 | **36** | | |
| 125-250 µm | 8.3 | | | 9.61 | **64** | | | 10.29 | **79** | 10.79 | **98** | 11.27 | **100** | 11.98 | **56** | | | 13.79 | **31** | | | 16.75 | **40** | | |
| Impact melt lithology | | | | | | | | | | | | | | | | | | | | | | | | | |
| 0-25 µm | 8.7 | | | 9.71 | **74** | | | 10.25 | **74** | | | 11.26 | **100** | | | 13.05 | **91** | | | 15.99 | **57** | | | 17.29 | **50** |
| 25-63 µm | 8.48 | | | 9.74 | **65** | | | 10.27 | **75** | | | 11.23 | **100** | 11.98 | **53** | | | | | | | 16.32 | **36** | | |
| 63-125 µm | 8.51 | | | 9.75 | **64** | | | 10.28 | **75** | | | 11.23 | **100** | 11.98 | **52** | | | | | | | 16.34 | **33** | | |
| 125-250 µm | 8.38 | | | 9.74 | **64** | | | 10.27 | **73** | | | 11.23 | **100** | 11.98 | **53** | | | | | | | 16.39 | **33** | | |

Table A1a: Band position bulk powder FTIR studies. Band positions in µm. Bold number right of RB: Intensity of feature in percent (%) normalized of strongest band

| Moderately shocked light lithology | | | | | | |
|---|---|---|---|---|---|---|
| 0-25 µm | 2.97 | 3.42 | 3.50 | 5.66 | 6.02 | 8.80 |
| 25-63 µm | 2.95 | 3.41 | 3.50 | 5.66 | 6.02 | 8.64 |
| 63-125 µm | 2.97 | 3.41 | 3.50 | 5.65 | 6.02 | 8.57 |
| 125-250 µm | 2.96 | 3.41 | 3.50 | 5.66 | 6.02 | 8.49 |
| Shock darkened lithology | | | | | | |
| 0-25 µm | | | | 5.66 | 6.02 | 8.77 |
| 25-63 µm | | | | 5.66 | 6.02 | 8.50 |
| 63-125 µm | | | | 5.66 | 6.02 | 8.47 |
| 125-250 µm | | | | 5.66 | 6.02 | 8.3 |
| Impact melt lithology | | | | | | |
| 0-25 µm | 2.9 | 3.42 | 3.5 | 5.65 | 6.02 | 8.7 |
| 25-63 µm | 2.8 | 3.42 | 3.5 | 5.65 | 6.02 | 8.48 |
| 63-125 µm | 2.8 | 3.41 | 3.5 | 5.65 | 6.02 | 8.51 |
| 125-250 µm | 2.8 | 3.4 | 3.49 | 5.65 | 6.02 | 8.38 |

Table A1b: Water and volatile features in the bulk powder spectra. Band position in µm.

| | CF | | | | | | | | | | | | | | | | | | | | | | | | | |
|---|---|---|---|---|---|---|---|---|---|---|---|---|---|---|---|---|---|---|---|---|---|---|---|---|---|---|
| Moderately shocked, light lithology | | | | | | | | | | | | | | | | | | | | | | | | | | |
| Chel2_1 | 8.55 | | | 9.57 | 59 | | | 10.31 | 87 | | | 10.77 | 100 | 11.31 | 97 | 11.95 | 52 | | | | | | | 13.79 | 21 | 14.57 | 17 |
| Chel2_2 | 8.5 | 9.23 | 79 | 9.57 | 69 | | | 10.24 | 77 | 10.5 | 69 | 10.78 | 77 | 11.3 | 100 | | | | | | | 13.48 | 17 | 13.81 | 25 | 14.58 | 21 |
| Chel2_3 | 8.6 | 9.17 | 17 | | | 9.97 | 72 | 10.33 | 87 | | | 10.69 | 100 | 11.26 | 98 | 11.97 | 46 | | | | | | | 13.79 | 19 | 14.53 | 12 |
| Chel2_4 | 8.55 | | | 9.58 | 65 | 9.96 | 44 | 10.33 | 84 | | | 10.74 | 100 | 11.29 | 87 | 11.99 | 47 | 12.6 | 30 | | | | | | | 14.55 | |
| Chel4_1 | 8.19 | | | 9.76 | 45 | | | 10.33 | 68 | | | 10.87 | 100 | 11.15 | 100 | 11.98 | 50 | | | | | | | | | | |
| Chel4_3 | 8.5 | 9.16 | 100 | | | 9.98 | 41 | 10.4 | 61 | | | 10.81 | 59 | 11.26 | 61 | | | | | | | 13.49 | 17 | 13.81 | 21 | | |
| Chel4_4 | 8.17 | 9.22 | 99 | 9.56 | 94 | | | 10.28 | 98 | 10.45 | 96 | 10.77 | 95 | 11.31 | 100 | | | 12.33 | 44 | 12.99 | 33 | 13.48 | 22 | 13.81 | 27 | 14.59 | 24 |
| Chel5_3 | 8.69 | 9.17 | 27 | 9.58 | 36 | | | 10.33 | 57 | | | 10.85 | 95 | 11.24 | 100 | 11.99 | 53 | | | | | | | 13.8 | 22 | 14.5 | 16 |
| Impact melt-like | | | | | | | | | | | | | | | | | | | | | | | | | | | |
| Chel4_2 | 8.73 | | | 9.69 | 53 | | | 10.33 | 56 | | | 10.75 | 100 | | | 12 | 41 | | | | | | | | | | |
| Chel5_1 | 8.7 | | | 9.6 | 37 | | | 10.33 | 53 | | | 10.86 | 100 | | | 11.99 | 45 | | | | | 13.13 | 22 | 13.56 | 17 | | |
| Chel5_2 | 8.56 | 9.21 | 10 | 9.58 | 31 | | | 10.36 | 46 | | | 10.85 | 100 | | | 11.99 | 39 | 12.51 | 24 | 12.97 | 19 | 13.39 | 15 | 13.81 | 13 | | |
| Chel2_5 | 9.8 | | | | | | | 10.3 | 39 | | | | | 11.15 | 100 | 11.98 | 39 | | | | | | | | | | |

Table A2a

| Shock darkened lithology | CF | | | | | | | | | | | | | | | | | | | | |
|---|---|---|---|---|---|---|---|---|---|---|---|---|---|---|---|---|---|---|---|---|---|
| Chel1_5 | 8.1 | 9.2 | **81** | 9.59 | **82** | 9.74 | **82** | 10.33 | **97** | 10.79 | **100** | 11.25 | **98** | 11.97 | **61** | | | 13.86 | **35** | | |
| Chel1_6 | 8.7 | | | | | | | 10.28 | **86** | 10.77 | **96** | 11.27 | **100** | 11.96 | **56** | | | 13.81 | **27** | 14.58 | **24** |
| Chel6_7 | 8.22 | 9.26 | **63** | | | | | 10.33 | **83** | 10.84 | **96** | 11.29 | **100** | | | | | 13.79 | **30** | 14.8 | **25** |
| Chel6_8 | 8.71 | | | 9.69 | **52** | 9.91 | **54** | 10.31 | **85** | 10.69 | **99** | 11.3 | **100** | 11.99 | **58** | | | | | 14.62 | **18** |
| Chel7_1 | 8.64 | | | 9.64 | **38** | | | 10.32 | **66** | 10.82 | **100** | 11.24 | **98** | 11.98 | **55** | | | 13.75 | **23** | | |
| Chel7_3 | 8.16 | 9.15 | **94** | | | 9.98 | **61** | 10.45 | **90** | 10.79 | **95** | 11.28 | **100** | | | 13.48 | **29** | 13.81 | **32** | 14.54 | **23** |
| | | | | | | | | | | | | | | | | | | | | | |
| Impact melt-like | | | | | | | | | | | | | | | | | | | | | |
| Chel1_3 | 8.27 | | | 9.85 | | | | 10.3 | **80** | | | 11.23 | **100** | 11.99 | **58** | | | | | | |
| Chel1_4 | 8.34 | | | 9.68 | | | | 10.31 | **60** | | | 11.22 | **100** | 11.98 | **60** | | | | | | |
| Chel4_5 | 8.71 | | | 9.95 | | | | 10.32 | **63** | | | 11.24 | **100** | 11.97 | **56** | | | | | | |
| Chel4_6 | 8.12 | | | 9.79 | | | | 10.31 | **84** | | | 11.17 | **100** | 11.99 | **61** | | | | | | |
| Chel5_4 | 8.14 | | | 9.65 | | | | 10.32 | **62** | 10.86 | **100** | | | 11.99 | **49** | | | | | | |
| Chel5_5 | 8.66 | | | 9.62 | | | | 10.32 | **71** | 10.63 | **81** | 11.3 | **100** | 11.99 | **62** | | | 13.66 | **27** | | |
| Chel7_2 | 8.13 | | | 9.5 | | | | 10.32 | **53** | | | 11.24 | **100** | 11.98 | **53** | | | | | | |
| Chel6_6 | 9.13 | | | 9.66 | | | | 10.32 | **48** | | | 11 | **100** | 11.96 | **41** | | | | | | |

Table A2b

| Impact melt lithology | CF | | | | | | | | | | | | | | |
|---|---|---|---|---|---|---|---|---|---|---|---|---|---|---|---|
| Chel1_1 | 8.06 | 9.78 | **35** | 10.31 | **55** | 11.24 | **100** | 11.98 | **56** | | | | | | |
| Chel1_2 | 8.06 | 9.8 | **39** | 10.31 | **59** | 11.23 | **100** | 11.98 | **59** | | | | | | |
| Chel6_2 | 8.08 | 9.68 | **39** | 10.31 | **61** | 11.2 | **100** | 11.98 | **58** | | | | | | |
| Chel6_3 | 8.06 | 9.74 | **42** | 10.31 | **60** | 11.22 | **100** | 11.98 | **58** | | | | | | |
| Chel6_5 | 8.01 | 9.76 | **35** | 10.3 | **59** | 11.26 | **100** | 11.99 | **60** | | | | | | |
| 'Pure' Impact melt lithology | | | | | | | | | | | | | | | |
| Chel4_7 | 8.03 | 9.78 | **33** | 10.3 | **50** | 11.26 | **100** | 11.98 | **57** | | | | | | |
| Chel4_8 | 8.04 | 9.75 | **37** | 10.3 | **53** | 11.28 | **100** | 11.98 | **57** | | | 13.87 | **23** | 14.86 | **14** |
| Chel4_9 | 8.02 | 9.77 | **27** | 10.3 | **43** | 11.3 | **100** | 11.98 | **54** | 13.13 | **28** | | | | |
| Chel4_10 | 8.03 | 9.77 | **35** | 10.3 | **53** | 11.29 | **100** | 11.98 | **57** | | | | | | |
| Chel6_1 | 8.08 | | | 10.3 | **54** | 11.29 | **100** | 11.98 | **57** | | | | | | |
| Chel6_4 | 8.1 | 9.77 | **26** | 10.3 | **49** | 11.24 | **100** | 11.98 | **55** | | | | | | |

Table A2c

Table A2a-c: Band positions from micro-FTIR studies. Band positions in µm. Bold number right of RB: Intensity of feature in percent (%) normalized to the strongest band.

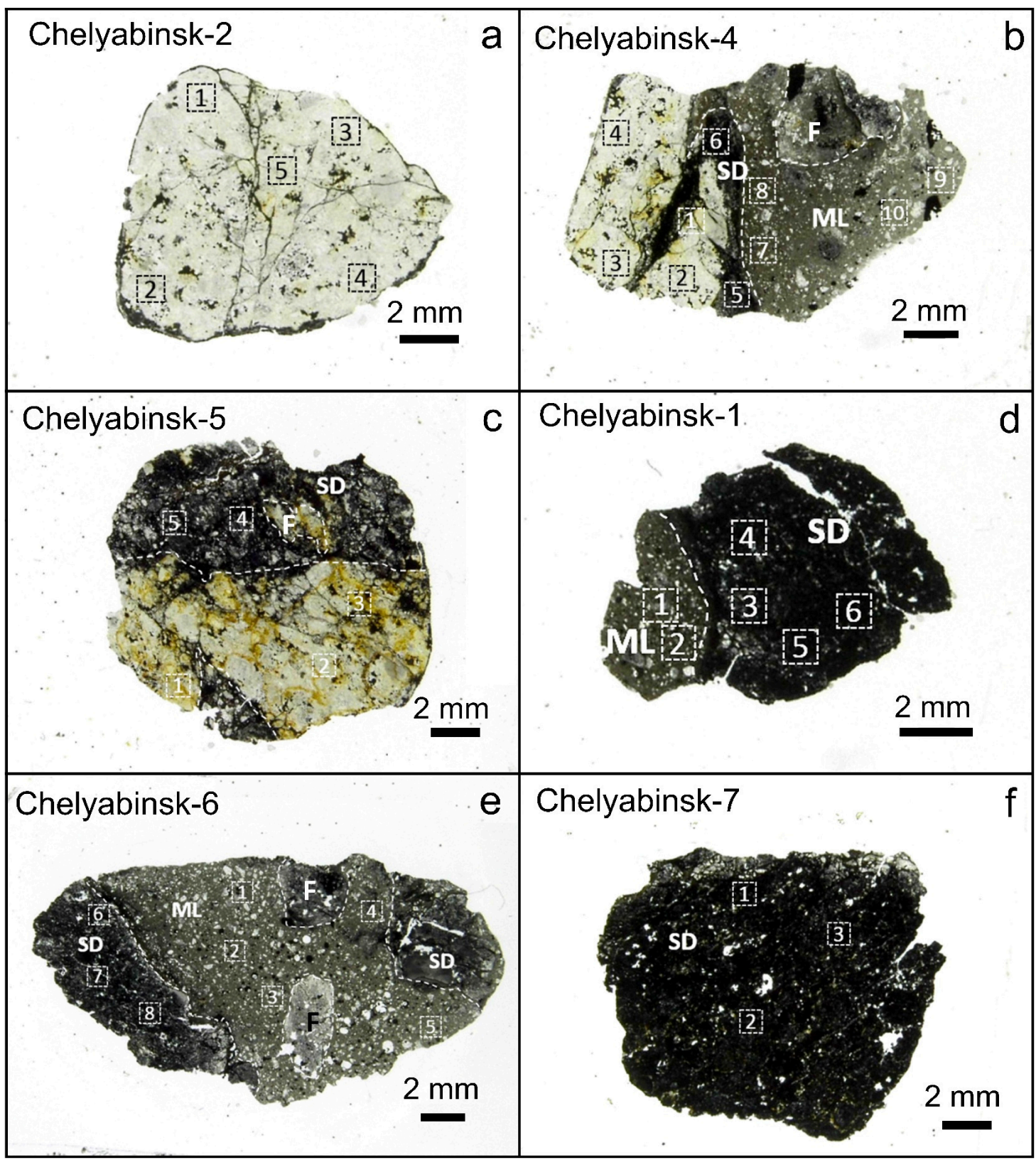

Fig.1

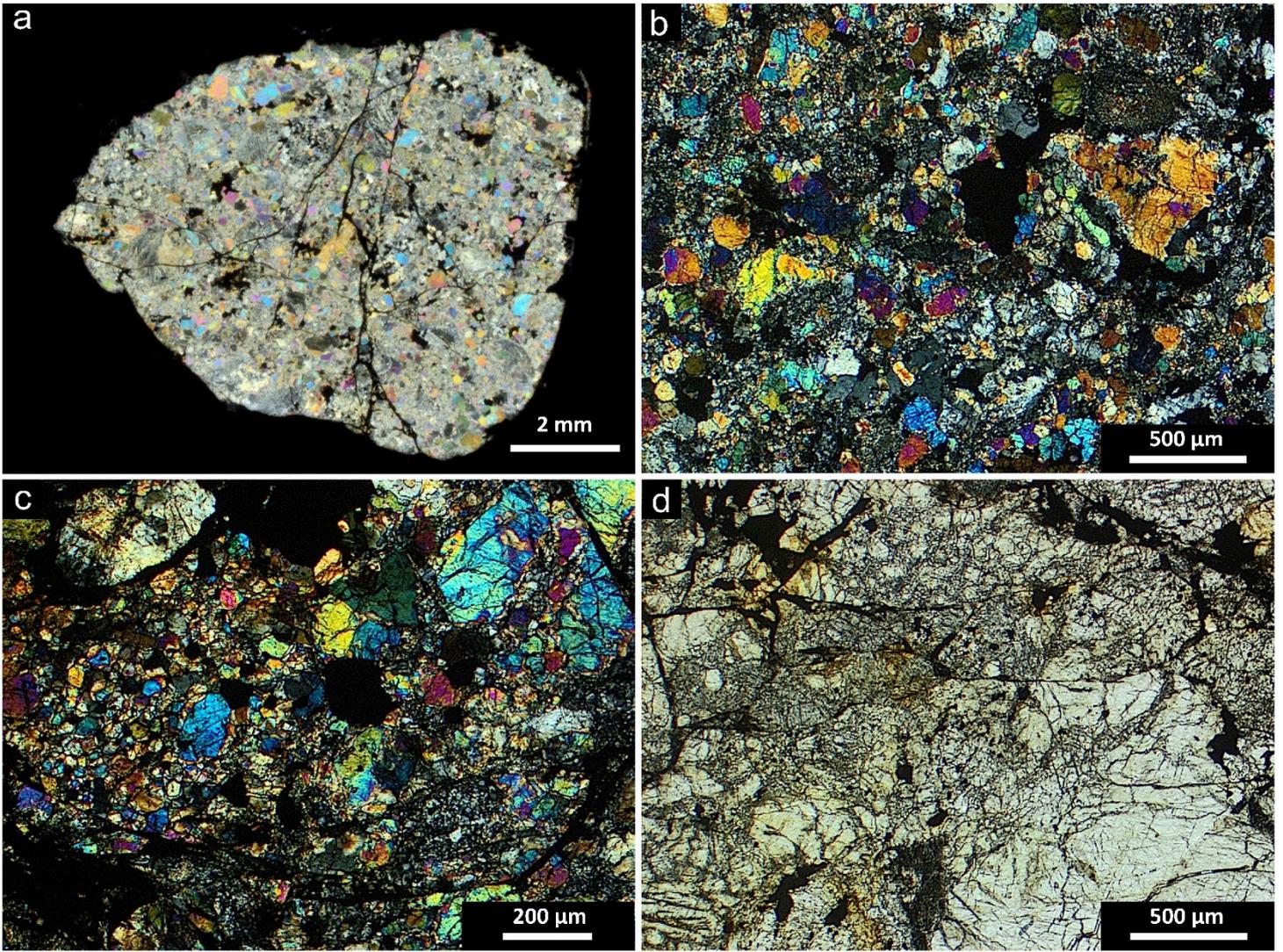

Fig.2

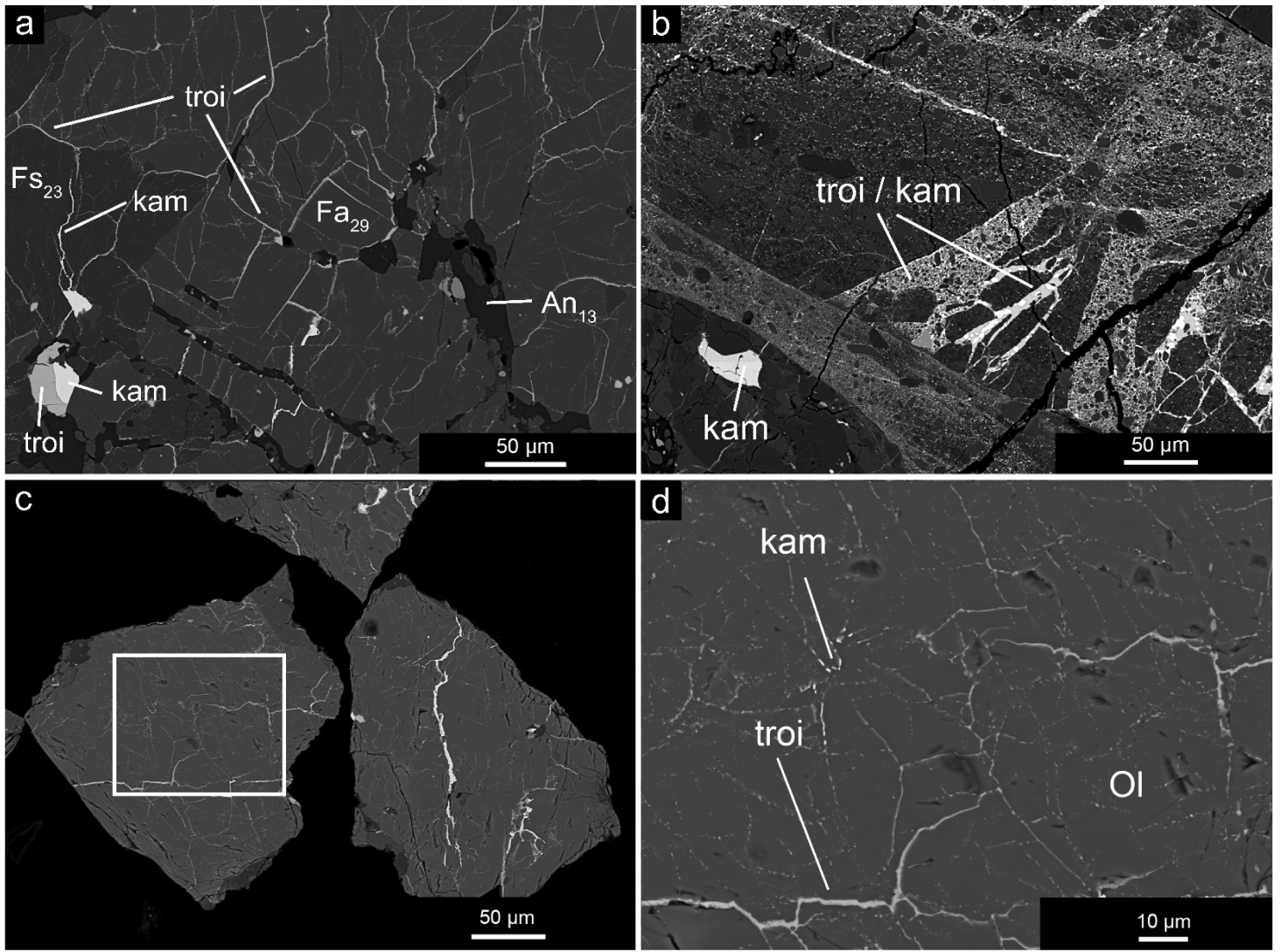

Fig.3

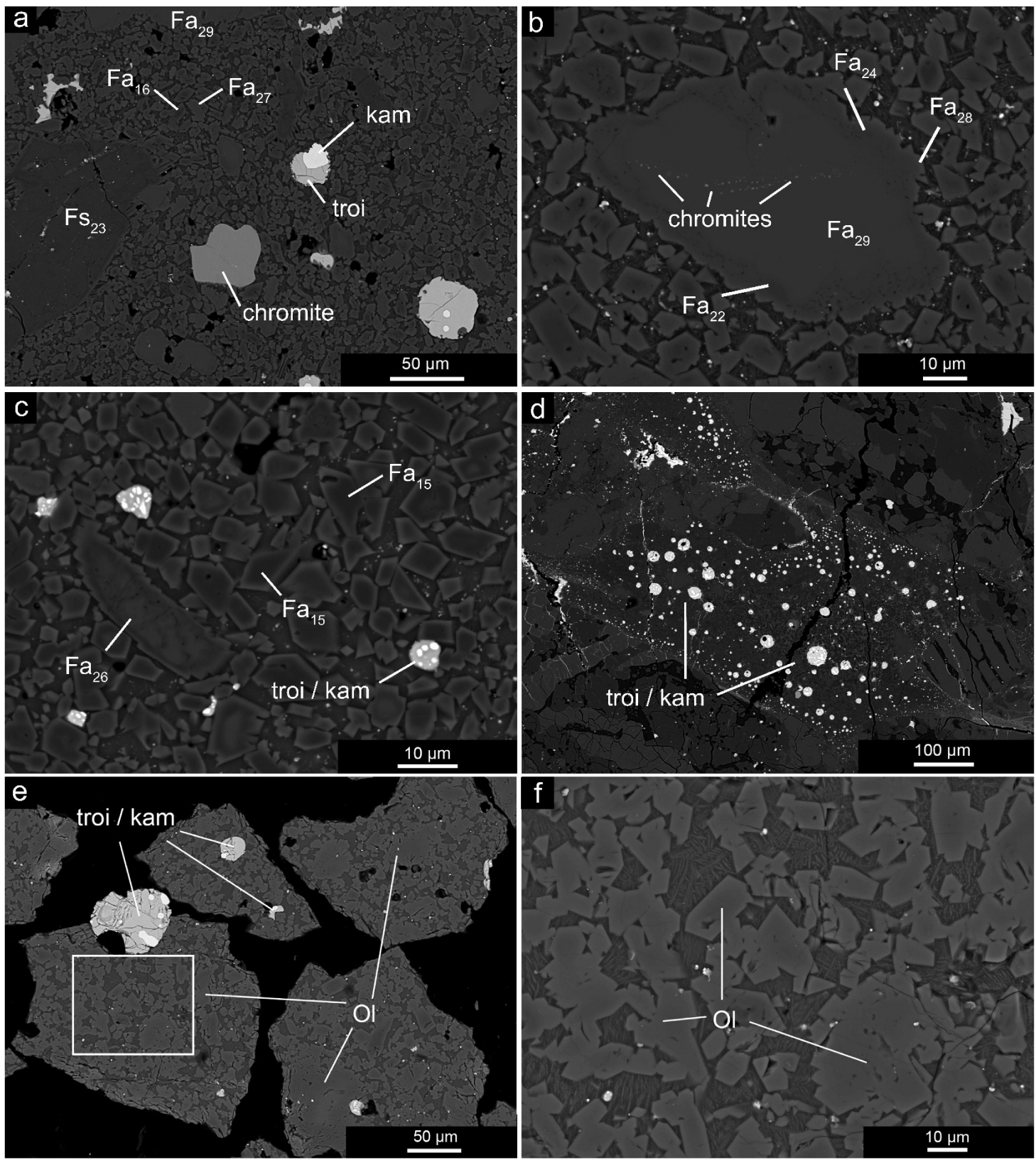

Fig.4

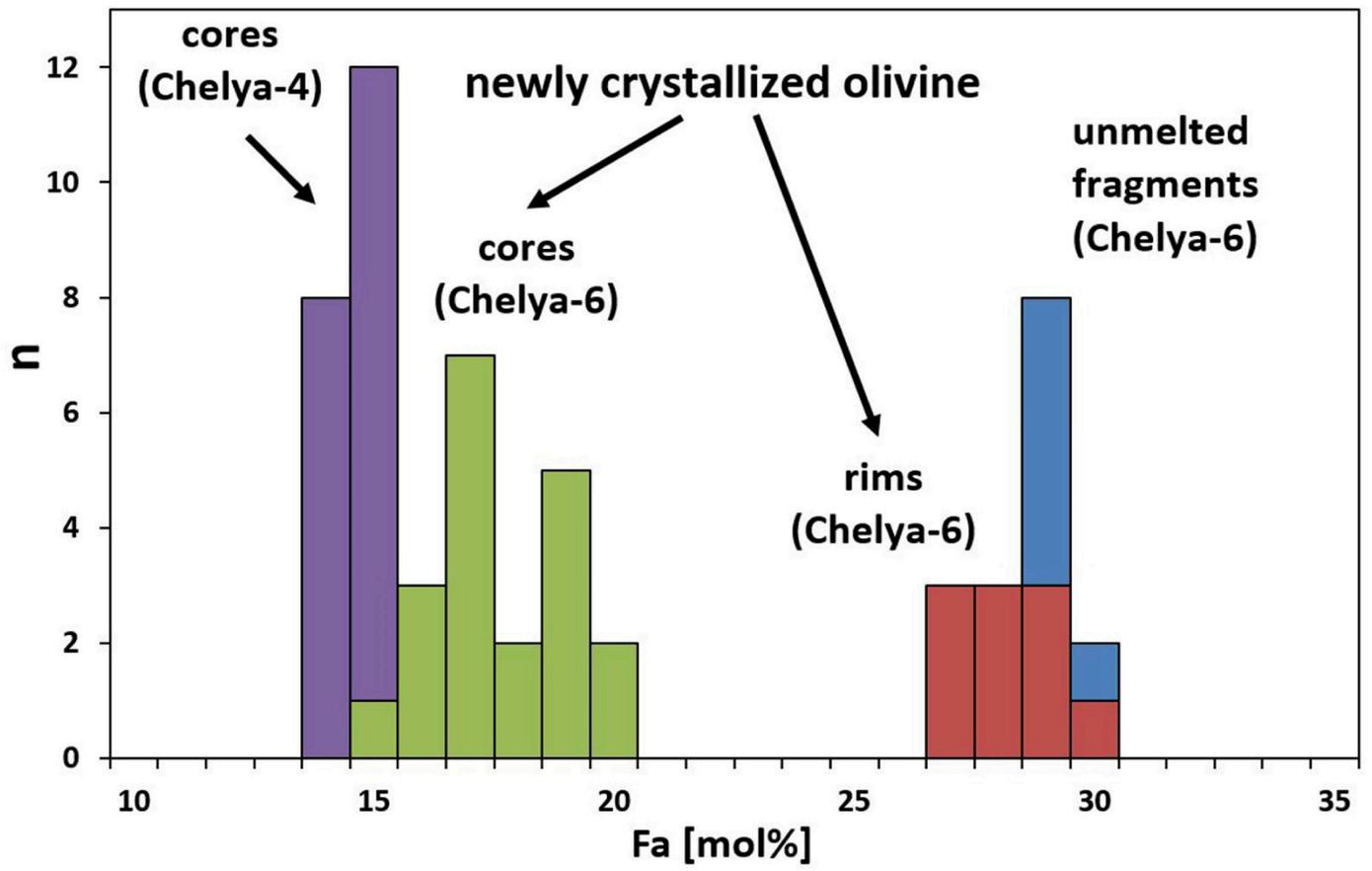

Fig.5

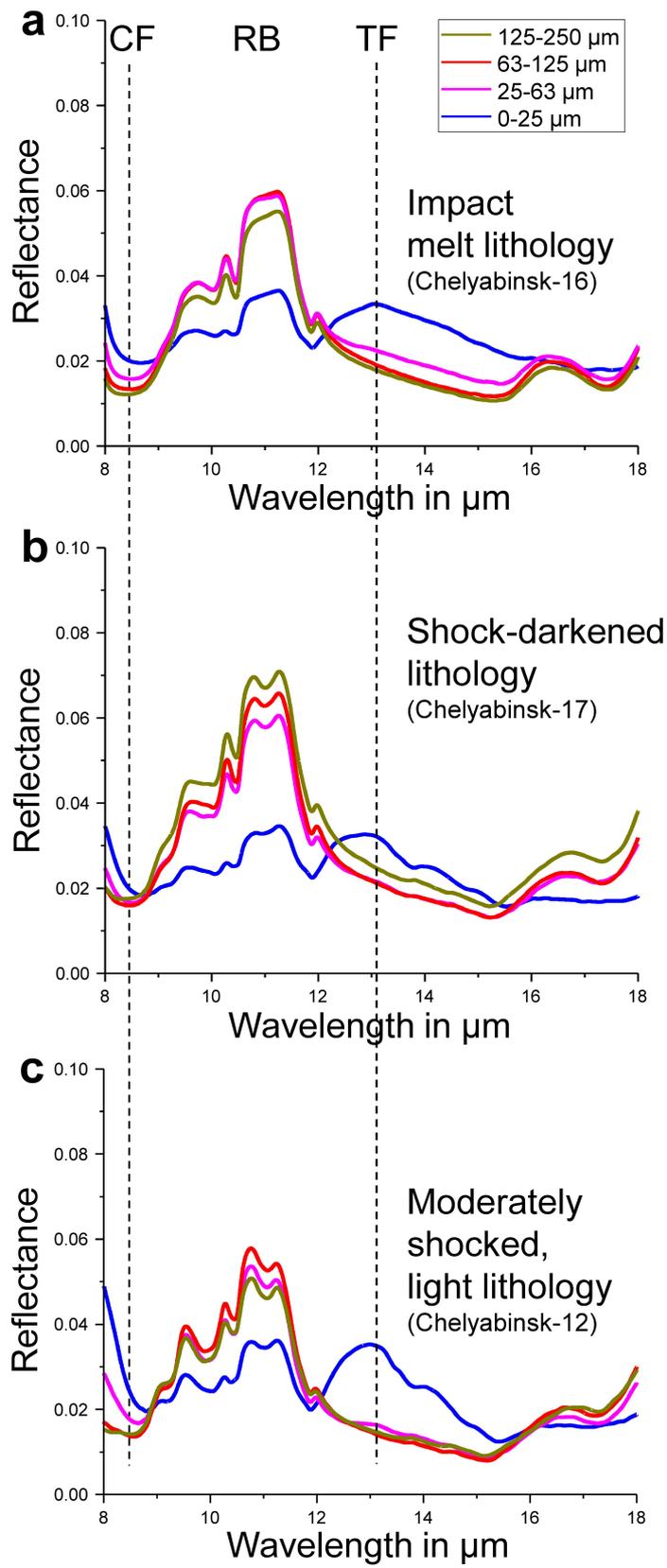

Fig.6

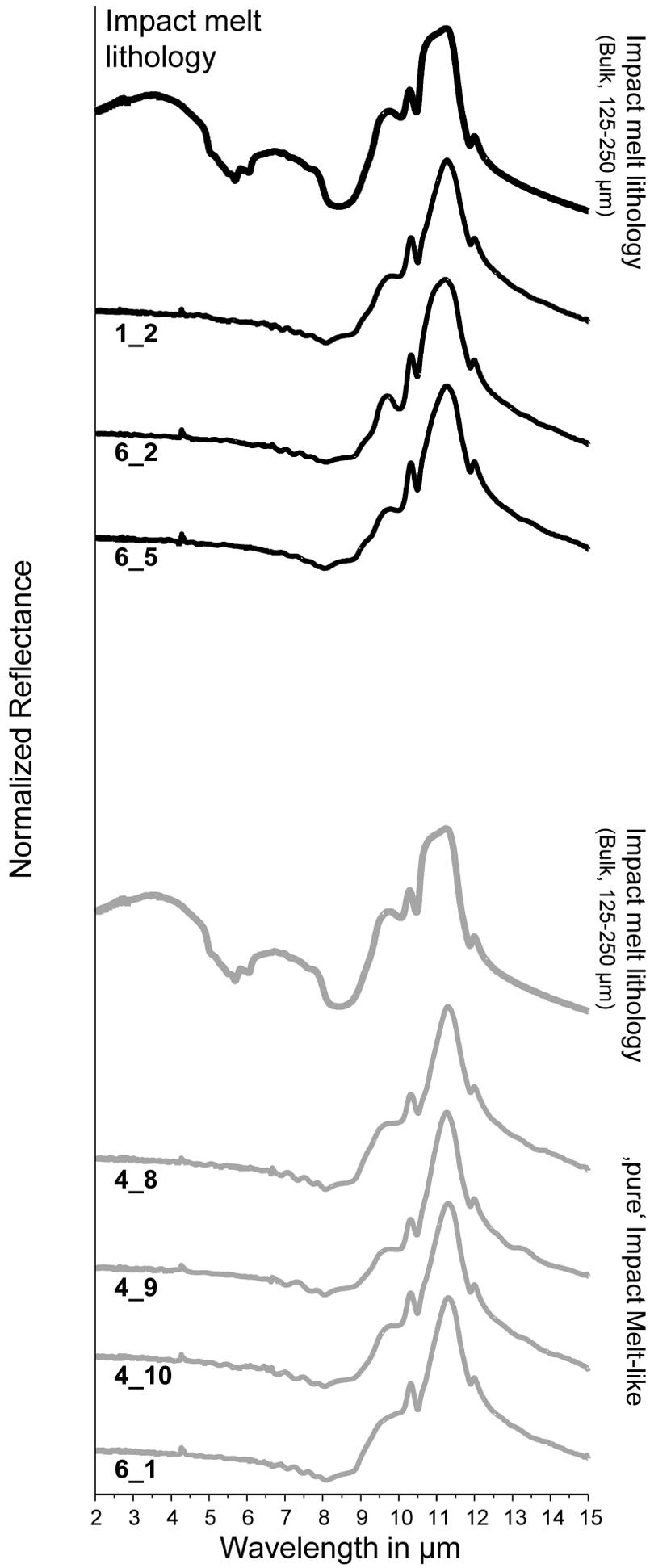

Fig.7a

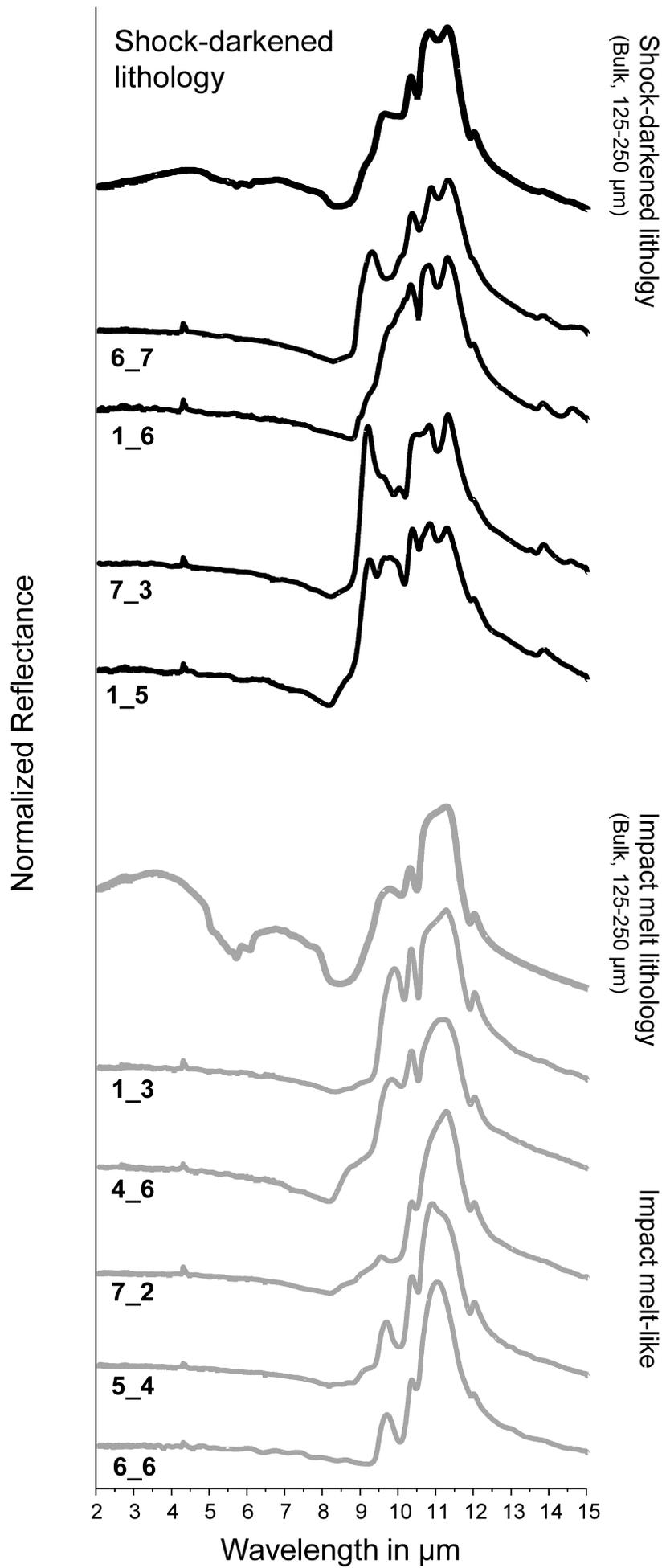

Fig.7b

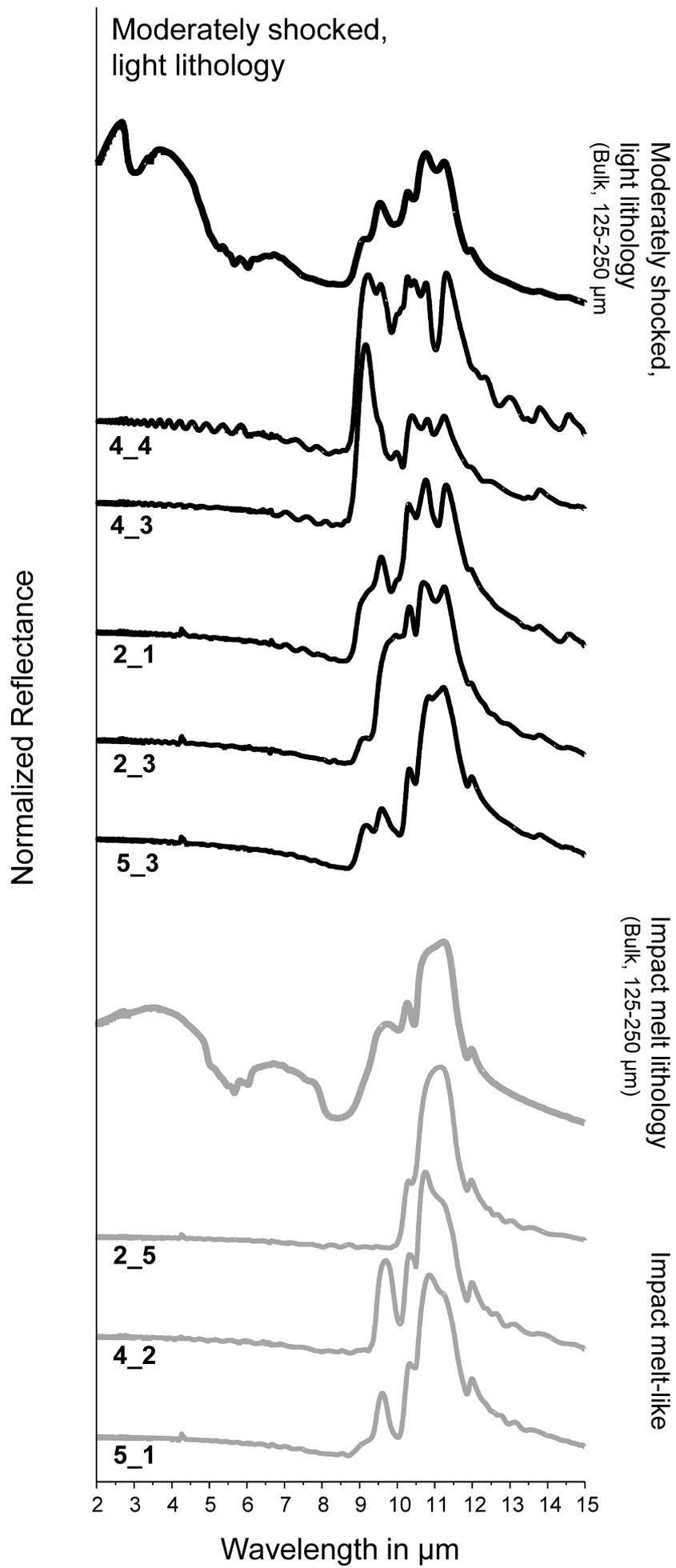

Fig.7c

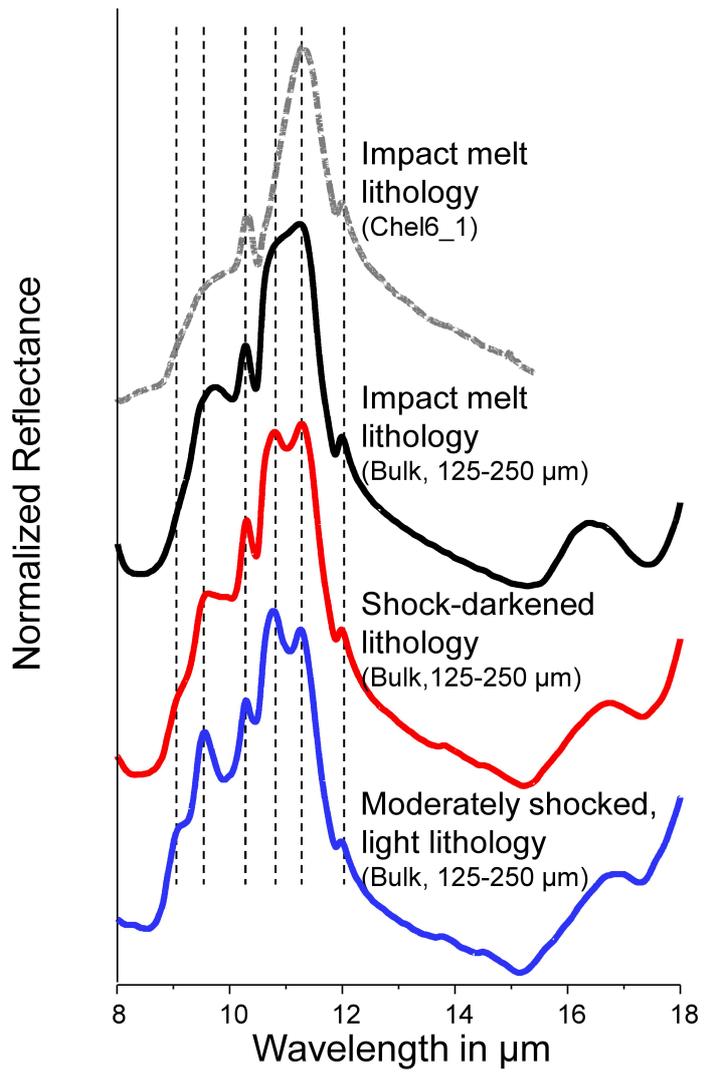

Fig.8

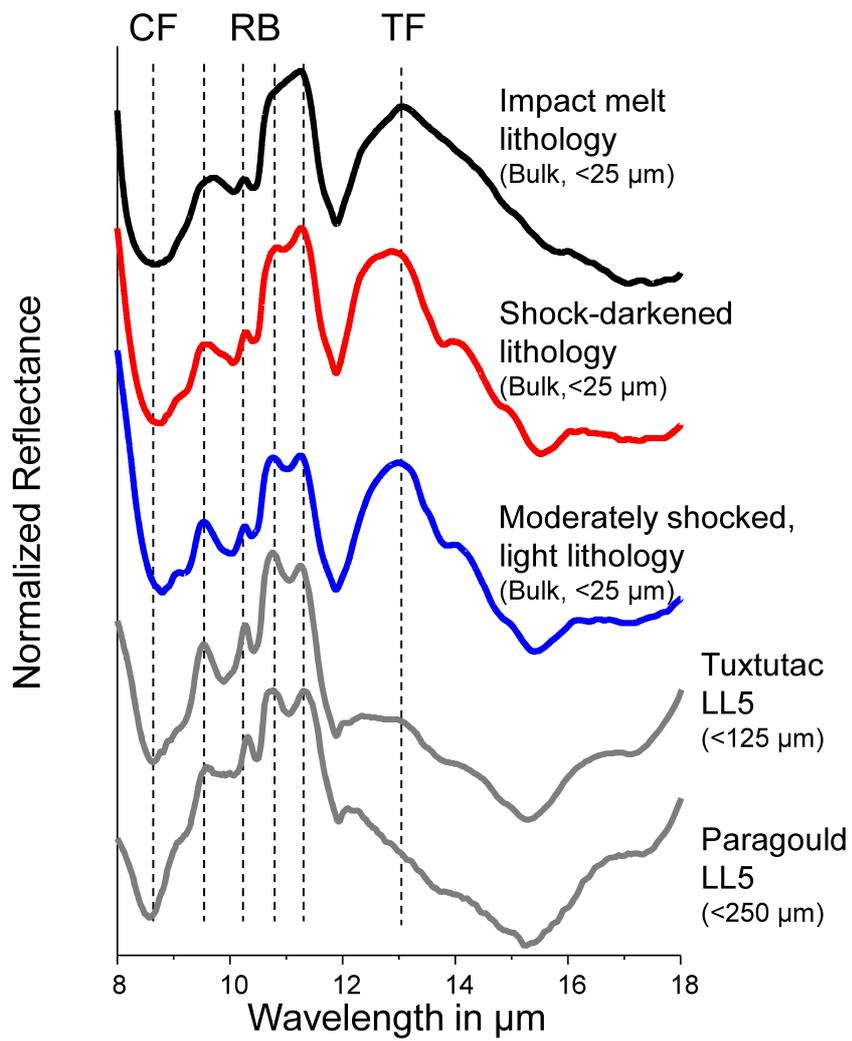

Fig.9